\documentclass[aps,twocolumn,preprintnumbers,nofootinbib,prd,superscriptaddress,10pt]{revtex4-1}

\usepackage[utf8]{inputenc}

\usepackage{epsf}  
\usepackage{graphicx}
\usepackage{amsmath}
\usepackage{amsfonts,color}
\usepackage{amssymb,float}
\usepackage{ulem}
\usepackage{booktabs}
\usepackage{mathrsfs}
\usepackage[caption=false]{subfig}
\usepackage{tikz}
\usetikzlibrary{shapes.misc, positioning}
\usepackage{microtype}
\usepackage[english]{babel}
\usepackage{blindtext}

\begin{document}
\title{Universal gravitational self-force for a point mass orbiting around a compact star}

\author{Xuefeng Feng}
\email{fengxuefeng@bimsa.cn}
\affiliation{Beijing Institute of Mathematical Sciences and Applications, Yanqihu, Huairou District, Beijing, 101408, China}
\affiliation{Yau Mathematical Sciences Center, Tsinghua University, Beijing, 100084, China}

\author{Huan Yang}
\email{hyangdoa@tsinghua.edu.cn}
\affiliation{Department of Astronomy, Tsinghua University, Beijing 100084, China}
\affiliation{Perimeter Institute for Theoretical Physics, Waterloo, Ontario N2L 2Y5, Canada}
\affiliation{University of Guelph, Guelph, Ontario N1G 2W1, Canada}
\begin{abstract}
In this work, we study the gravitational back-reaction (i.e., the ``self-force") of a point mass moving around a non-rotating, compact star on a circular orbit. We find that the additional self-force, comparing with the case with a point mass orbiting around a Schwarzschild black hole, can be well characterized by an universal frequency-dependent function multiplied by the (dynamical) tidal deformability of the compact star. This finding provides the foundation for building the waveform model for an extreme mass-ratio inspiral system around a star-like black hole mimicker, which is relevant for testing General Relativity and exotic compact objects with space-borne gravitational wave detectors. 
\end{abstract}
\maketitle
\section{Introduction}
Extreme mass-ratio inspirals (EMRIs) are one of the two primary extragalactic sources of space borne gravitational wave detectors, such as LISA (Laser Interferometer Space Antenna), Tianqin and Taiji \cite{Baker:2019nia,TianQin:2020hid,Hu:2017mde}. The dominant formation channels include  scattering process in nuclear star clusters \cite{Babak:2017tow} and migration of stellar-mass black holes in Active Galactic Nuclei according to the rate calculations \cite{Levin:2006uc,Pan:2021ksp,Pan:2021oob}. There are recent rate estimates about accelerated EMRI formation around supermassive black hole binaries \cite{Naoz:2022rru}, but the rate is still rather uncertain \cite{Bode:2013mma,Mazzolari:2022cho} because of the unknown population of close supermassive black hole binaries, the efficiency of sustaining the Kozai-Lidov mechanism and replenishing stellar-mass black hole supplies in the nuclear cluster. It is also interesting to characterize the EMRI properties (e.g. mass and spins \cite{Pan:2021xhv}) in various formation channels to help distinguish them in future detection.

Because of the large number of cycles ($10^4-10^5$) in the detection band, EMRIs are very sensitive probes of small environmental forces \cite{Bonga:2019ycj,Yunes:2011ws,Barausse:2014pra}, possible superadiantly excited Axion clouds \cite{Zhang:2019eid,Zhang:2018kib,Hannuksela:2018izj}, multipole moments of the background spacetime \cite{Ryan:1995wh,Raposo:2018xkf,Tahura:2023qqt} and the black hole nature of the central massive object (see the review article \cite{Cardoso:2019rvt} for various classes of black hole mimickers). These are all important science goals of space borne gravitational wave missions \cite{Barausse:2020rsu,LISA:2022kgy,LISAConsortiumWaveformWorkingGroup:2023arg}. In particular, in order to test the nature of the central object it is necessary to build a more complete description for the dynamics and waveform of a stellar-mass object moving around a massive black hole mimicker. Because of the large number of black hole mimicker options, it is also beneficial if the waveform of mimickers only (approximately) depends on a few parameters, e.g. the tidal deformability and ``horizon reflectivity" \cite{Maggio:2021uge}, without using fine structure or composition details of the mimickers. This is also one of the primary motivations of this work.

We consider a model problem with a point mass moving along a circular trajectory around a central compact star. The unperturbed configuration of the star is constructed using the Tolman–Oppenheimer–Volkoff (TOV) equation with a set of equation of state. With the presence of the point mass, the exterior metric perturbation (outside the star) is solved using the Regge-Wheeler-Zerilli formalism and the inner matter and metric perturbations are solved following similar exercise in  \cite{thorne1967non,lindblom1983quadrupole,kojima1987stellar,tominaga1999gravitational,gittins2020tidal}. The metric perturbations are matched on the star surface.

With the metric perturbations we are able to compute gravitational wave flux radiated at infinity. In particular, we are interested in comparing this flux to the one generated by a point mass circulating around a Schwarzschild black hole (the masses are the same) with the same orbital frequency. The flux deviation is dominated by the $\ell=2, m=2$ component, as expected, which is well fitted by the dynamical tidal deformability of the star times a function of orbital frequency. The higher order mode contributes roughly below a few percent of the tidal flux of the $\ell=m=2$ mode, except near the resonant frequency of some particular modes. We have chosen several polytropic equation of state for the star and find this tidal flux relation approximately universal among different EOS. Interestingly as we extrapolate the flux deviation in the limit that the tidal Love number is zero, we find the flux from a compact star is indistinguishable from the infinity flux of the black hole case within the numerical precision of our implementation. Similar observations were made in \cite{datta2020tidal,fransen2021modeling,maggio2021extreme} using {\it ad hoc} reflection conditions in a Teukolsky solver. 

With the metric perturbations we also compute the gravitational self force acting on the point mass. Regularization is not necessary in this case as we are interested in the difference of self force between the compact star scenario and the black hole scenario. We find that the $t$-component of the self force is equal to the flux radiated at infinity and the $\phi$-component of the self force can be obtained by using $\Omega F^\phi =m F^t$, where $m$ is the azimuthal number and $\Omega$ is the orbital frequency. The $r$-component self force satisfies a rather linear relation with the dynamical tidal deformability, which is also universal for the EOSs we have checked. As a result, both the conservative and dissipative part of the self force, as deviated from the black hole case, can be characterized by the dynamical tidal deformability of the central body and a function of orbital frequency that we fit through our numerical calculations. We can then construct an EMRI waveform model for general star-like black hole mimickers that contains two free parameters for their finite-size effects: the tidal deformability and the f-mode frequency. It is actually interesting to explore other types of mimickers, such as a boson star, to see whether the universal relation found here still applies. Indeed for a black hole mimicker with size smaller than the radius of Inner Most Stable Orbit, we show that the tide-induced phase modulation can reach the level of $\mathcal{O}(10^2) - \mathcal{O}(10^3)$ rad for a typical EMRI around a $10^6$ $M_\odot$ massive black hole. Therefore such a waveform model is crucial for the search and identification of black hole mimickers using EMRI systems.

This article is organized as follows. In Section II, we present an introduction to the equilibrium stellar model obtained by solving the Tolman-Oppenheimer-Volkoff equations. Section III is dedicated to deriving the exterior and interior perturbation equations for both metric and fluid perturbations. In Section IV, we explicitly compute the tide-induced gravitational wave flux and derive the universal gravitational self-force for a compact star with an orbiting point mass. Finally, we summarize in Section V. Throughout this paper except in Sec.~\ref{sec:tct}, where physical units are used to construct stellar-mass stars, we adopt geometrical units, $G=c=1$, where $G$ denotes the gravitational constant and $c$ the speed of light, respectively.
\section{Equilibrium configuration}

\begin{figure} 
    \centering  \includegraphics[height=6cm,width=8.6cm]{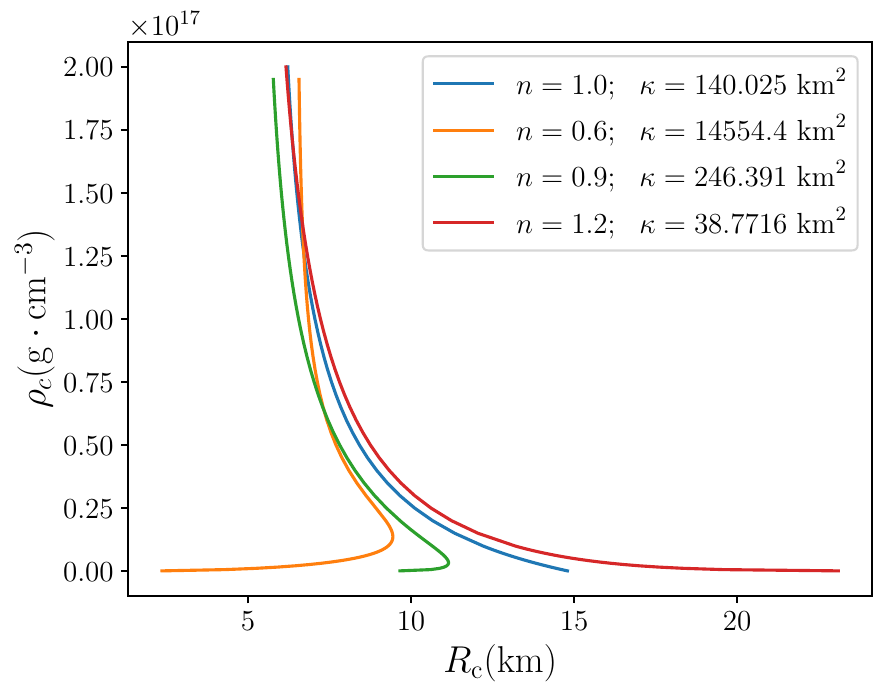}
    \caption{The relationship between the central density $\rho_{c}$ of a stellar-mass compact star and the radius $R_{c}$ of the star. These explicit equations of state of the stars are selected from Fig.~\ref{fig:flux1}.}
    \label{fig:radius}
\end{figure}

In this section, we briefly explain the construction of the equilibrium stellar model. The geometry of the star can be characterized by a static spherically symmetric metric, which takes the form \cite{thorne1967non}
\begin{align}\label{eq:sph_metric}
	ds_{0}^2=-e^{\nu}dt^2+e^{\lambda}dr^2+r^2(d\theta^2+\sin^2\theta d\varphi^2),
\end{align}
where 
\begin{align}
	e^{\lambda}=\left (1-\frac{2\mathcal{M}}{r} \right )^{-1}\nonumber.
\end{align}
The function $\nu$ depends solely on $r$, while the mass function $\mathcal{M}(r)$ represents the gravitational mass enclosed within a radius $r$. This mass function must vanish at the center, $\mathcal{M}(0)=0$, and the total mass of the star is $\mathcal{M}(R)=M$, where $R$ denotes the star's radius.
Outside the star, $r\geq R$, the metric becomes
\begin{align}
	e^{\nu}=e^{-\lambda}=1-\frac{2M}{r}.
\end{align}

The state of hydrostatic equilibrium is governed by the Tolman-Oppenheimer-Volkoff equations \cite{oppenheimer1939massive}.
\begin{align}
	\frac{dp}{dr}&=-\frac{(\rho+p)(\mathcal{M}+4\pi r^3 p)e^{\lambda}}{r^2},\label{eq:tov1}\\
	\frac{d\nu}{dr}&=\frac{2(\mathcal{M}+4\pi r^3 p)e^{\lambda}}{r^2},\label{eq:tov2}\\
	\frac{d \mathcal{M}}{dr}&=4\pi r^2 \rho,\label{eq:tov3}
\end{align}
where $\rho$ and $p$ are the total energy density and pressure respectively. In this work,  we use a set of polytropic equations of state as follows:
\begin{align}\label{eq:eos}
	p=\kappa \rho^{1+1/n},
\end{align}
where $\kappa$ is the adiabatic constant and $n$ is polytropic index. In Fig.~\ref{fig:radius} we show a set of star configurations with star mass in the solar-mass range (relevant for ground-baesd detectors and the studies in \cite{Feng:2021sax}) and parameters used for Fig.~\ref{fig:flux1}. For EMRI-type of problems the mass will be orders of magnitude larger and the central density will be much smaller (see the related construction in Sec.~\ref{sec:tct}).

The adiabatic index $\gamma$, which dictates the response of stellar material to pulsational compressions, is expressed as
\begin{align}\label{eq:gamma}
	\gamma=\frac{\rho+p}{p}\frac{dp}{d\rho}
\end{align}
This index will be used in Sec.~\ref{sec:pin}.
\section{Gravitational perturbation}\label{sec:grap}

\begin{figure*}
    \includegraphics[height=10.0cm,width=17.5cm]{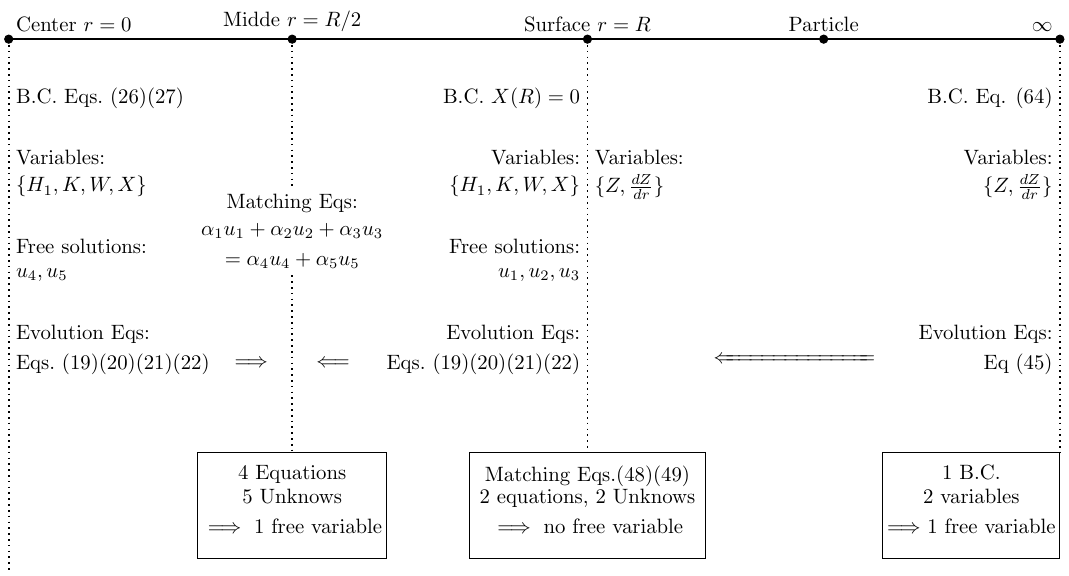}
    \caption{An illustration plot that shows how even-parity perturbations are solved with corresponding wave equations and boundary conditions. Here the relevant perturbative quantities inside the star are $u(r)=\{H_1(r),K(r),W(r),X(r)\}$. To more efficiently solve the system we further divide the star interior into two domains: $r \le R/2$ and $R/2 \le r \le R$. At the  surface of a compact star, we require that $X(R)=0$ so that $H_1(R)$, $K(R)$ and $W(R)$ can be chosen freely. This means that there are three linearly independent solutions $u_1(r)$, $u_2(r)$ and $u_3(r)$ with coefficients to be determined in the domain $R/2 \le r\leq R$. Note that inside the star the evolution equations are Eqs.~\eqref{eq:H1}, \eqref{eq:K}, \eqref{eq:W} and \eqref{eq:X}. At the center of a compact star, given regular boundary contions Eqs.~\eqref{eq:regular11} and \eqref{eq:regular12} we  have two independent solutions $u_4$ and $u_5$, which can be integrated outward using the evolution equations. On the $r=R/2$ surface the continuity of perturbative variables requires  $\alpha_1 u_1 +\alpha_2 u_2 + \alpha_3 u_3 =\alpha_4 u_4 + \alpha_5 u_5$. These four equations may be solved for the five independent constants $\alpha_{i}$, leave one free parameter to be determined by using the perturbations outside the star. In the star exterior there are in principle two free parameters for the Zerilli variables and its derivative. However, the outgoing condition at spatial infinity eliminates one free parameter, with the second free parameter determined by the matching conditions at the star suraface.
}
    \label{fig:Sketch}
\end{figure*}

Given the star equilibrium configuration and the background spacetime we now discuss the gravitational perturbations generated by an orbiting point mass. In principle the calculation can be divided into two separate regimes: the star interior and the star exterior. Within the star we need to account for both the fluid and gravitational perturbations, where outside the star only gravitational perturbations need to be considered. The basis strategy of solving the even-parity perturbation equations is outlined in Fig.~\ref{fig:Sketch}, with details presented in Sec.~\ref{sec:pin} (interior perturbation) and Sec.~\ref{sec:pout} (exterior perturbation) respectively.

\subsection{Perturbations inside the star}\label{sec:pin}
In order to determine the radiation generated by  a  massive point particle orbiting around a compact star, it is natural  to consider  perturbations  inside and outside the star separately. For the metric perturbations,   we follow  the formalism in \cite{thorne1967non,lindblom1983quadrupole,kojima1987stellar,tominaga1999gravitational,gittins2020tidal} in the Regge-Wheeler gauge
\begin{align}
    ds^2 = ds_{0}^2 + ds_{\rm odd}^2 + ds_{\rm even}^2,
\end{align}
where $ds_{0}^2$ is given by Eq.~\eqref{eq:sph_metric}, and the odd/even parity parts are
\begin{align}
ds_{\rm odd}^2&=2h_{0}\left(-\frac{1}{\sin\theta}\frac{\partial Y_{\ell m}}{\partial\varphi}dtd\theta+\sin\theta\frac{\partial Y_{\ell m}}{\partial\theta}dtd\varphi\right)e^{-i\omega t}\nonumber\\
&\quad+2h_{1}\left(-\frac{1}{\sin\theta}\frac{\partial Y_{\ell m}}{\partial\varphi}drd\theta+\sin\theta\frac{\partial Y_{\ell m}}{\partial\theta}drd\varphi\right)e^{-i\omega t}\,,\label{eq:metric_odd}\\
	ds_{\rm even}^2&=e^{\nu}\zeta^lH_{0}Y_{\ell m}e^{-i\omega t}dt^2+2i\omega r\zeta^{l}H_1 Y_{\ell m}e^{-i\omega t}dtdr\nonumber\\
	&\quad+\zeta^lH_0 Y_{\ell m}e^{-i\omega t}dr^2\nonumber\\
	&\quad+r^2\zeta^lKY_{\ell m}e^{-i\omega t}(d\theta^2+\sin^2\theta d\varphi^2)\,,\label{eq:metric_even}
\end{align}
with
\begin{align}
   \zeta= \begin{cases} r/R\quad 0\leq r \textless R\\ 1\qquad\  r\geq R, \end{cases} 
\end{align}
 the functions $H_0$, $H_1$ and $K$ are functions of only $r$ and $Y_{\ell m}(\theta,\varphi)$ is the spherical harmonics. 

The  metric perturbations give rise to the perturbation of the Einstein tensor $\delta G_{\mu\nu}$ according to
\begin{align}\label{eq:perturb}
	-2\delta G_{\mu\nu}&={h_{\mu\nu;\alpha}}^{;\alpha}-(f_{\mu;\nu}+f_{\nu;\mu})+2R^{\rho}{}_{\mu}{}^{\alpha}{}_{\nu}h_{\rho\alpha}\nonumber\\
	&\quad+h^{\alpha}{}_{\alpha;\mu;\nu}-(R^{\rho}{}_{\nu}h_{\mu\rho}+R^{\rho}{}_{\mu}h_{\nu\rho})\nonumber\\
	&\quad+g_{\mu\nu}(f_{\lambda}{}^{;\lambda}-h^{\alpha}{}_{\alpha;\lambda}{}^{;\lambda})+h_{\mu\nu}R\nonumber\\
	&\quad-g_{\mu\nu}h_{\alpha\beta}R^{\alpha\beta},\\
	f_{\mu}&=h_{\mu\alpha}{}^{;\alpha}.
\end{align}
where $R_{\mu\nu}$ and $R$ are the background Ricci curvature and scalar curvature. The Einstein's equation requires that $\delta G_{\mu\nu} =8 \pi \delta T_{\mu\nu}$, where the fluid perturbations of the star determines $\delta T_{\mu\nu}$. 

\subsubsection{Even parity}

The perturbations of the fluid variables in the stellar model are described with the fluid dislocation vector field $\xi^{a}$. In the appropriate gauge $\xi^{t}=0$, and the relevant components are given by
\begin{align}
	\xi^{r}&=\zeta^{\ell}r^{-1}e^{-\lambda/2}WY_{\ell m}e^{-i\omega t},\\
	\xi^{\theta}&=-\zeta^{\ell}r^{-2}V(\partial_{\theta}Y_{\ell m})e^{-i\omega t},\\
	\xi^{\varphi}&=-\frac{\zeta^{\ell}}{r^2\sin^2\theta}V(\partial_{\varphi}Y_{\ell m})e^{-i\omega t},
\end{align}
where $W$ and $V$ are functions of only $r$. Given the fluid displacement mentioned above, the fluid four-velocity can be given by \cite{nakamura1987general}
\begin{align}
    u_{\mu}&=(u_{t},u_{r},u_{\theta},u_{\phi})\nonumber\\
    &=\bigg(-e^{\nu/2}\big(1-\frac{1}{2}H_{0}Y_{\ell m}e^{-i\omega t}\big),\nonumber\\
    &\quad\quad e^{-\nu/2}\big(-\frac{i\omega}{r^2}W e^{\lambda/2}+H_{1}\big)Y_{\ell m}e^{-i\omega t},\nonumber\\
    &\quad\quad i\omega e^{-\nu/2}V(\partial_{\theta}Y_{\ell m})e^{-i\omega t},i\omega e^{-\nu/2}V(\partial_{\varphi}Y_{\ell m})e^{-i\omega t}\bigg).
\end{align}
Then the energy-momentum tensor of a perfect fluid can be written as
\begin{align}
    T_{\mu\nu}=(\rho+p)u_{\mu}u_{\nu}+pg_{\mu\nu}.
\end{align}

With the stress-energy perturbations of the fluid and after plugging the perturbed metrics Eq.~\eqref{eq:metric_even} into Eq.~\eqref{eq:perturb}, the system of equations can be obtained 
\begin{align}
	H_{1}'=&-\frac{1}{r}\left[\ell+1+\frac{2Me^{\lambda}}{r}+4\pi r^2 e^{\lambda}(p-\rho)\right]H_{1}\nonumber\\
	&+\frac{1}{r}e^{\lambda}[H_{0}+K-16\pi(\rho+p)V]\,,\label{eq:H1}\\
	K'=&\frac{1}{r}H_0+\frac{\ell(\ell+1)}{2r}H_1-\left[\frac{\ell+1}{r}-\frac{1}{2}\nu'\right]K\nonumber\\
	&-\frac{8\pi(\rho+p)e^{\lambda/2}}{r}W\,,\label{eq:K}\\
	W'=&-\frac{\ell+1}{r}W+re^{\lambda/2}\left[\frac{e^{-\nu/2}X}{\gamma p}-\frac{\ell(\ell+1)}{r^2}V+\frac{1}{2}H_0+K\right]\,,\label{eq:W}\\
	X'=&-\frac{\ell X}{r}+(\rho+p)e^{\nu/2}\biggl\{\frac{1}{2}\left(\frac{1}{r}-\frac{1}{2}\nu'\right)H_0-\frac{\ell(\ell+1)}{2r^2}\nu'V\nonumber\\
	&+\frac{1}{2}\left[r\omega^2e^{-\nu}+\frac{\ell(\ell+1)}{2r}\right]H_1+\frac{1}{2}\left(\frac{3}{2}\nu'-\frac{1}{r}\right)K\nonumber\\
	&-\frac{1}{r}\left[4\pi(\rho+p)e^{\lambda/2}+\omega^2e^{\lambda/2-\nu}-\frac{1}{2}r^2(\frac{e^{-\lambda/2}}{r^2}\nu')'\right]W\biggr\}\,,\label{eq:X}
\end{align}
where the functions $V$ and $H_0$ can be obtained by the following relations as a consequence of Einstein's equations
\begin{align}
	X&=\omega^2(\rho+p)e^{-\nu/2}V-\frac{p'e^{(\nu-\lambda)/2}}{r}W\nonumber
	\\
	&\quad+\frac{1}{2}(\rho+p)e^{\nu/2}H_0\,.
\end{align}
Here $X(R) = 0$ must be satisfied as the pressure must vanish at the star surfaces. 
In addition, $H_0, H_1, X, K$ are related by 
\begin{widetext}
	\begin{align}
	&\left[3M+\frac{1}{2}(\ell+2)(\ell-1)r+4\pi r^3p\right]H_0=8\pi r^3e^{-\nu/2}X-\left[\frac{1}{2}\ell(\ell+1)(M+4\pi r^3p)-\frac{\omega^2r^3}{e^{\lambda+\nu}}\right]H_1\nonumber\\
	&\quad\quad+\left[\frac{1}{2}(\ell+2)(\ell-1)r -\omega^2r^3e^{-\nu}+\frac{1}{r}e^{\lambda}(M+4\pi r^3p)(3M-r+4\pi r^3p)\right]K\,.
	\end{align}
\end{widetext}

The original set of perturbation equations was  derived by Thorne and Campolattaro in \cite{thorne1967non}. They were later obtained by many authors \cite{lindblom1983quadrupole,kojima1987stellar,detweiler1985nonradial,gittins2020tidal} in various conventions. It is important to note that equations Eq.~\eqref{eq:H1}, \eqref{eq:K}, \eqref{eq:W} and \eqref{eq:X} are singular near the center $r=0$ since $1/r$ is not analytic in a neighborhood of the center. As a result, we  assume that the solutions are represented by power series expansions near the center $r=0$, which are given by
\begin{align}\label{eq:zero}
	H_{1}(r)&=y_{0}+\frac{1}{2}y_{2}r^2+\cdots,\nonumber\\
	K(r)&=k_{0}+\frac{1}{2}k_{2}r^2+\cdots,\nonumber\\
	W(r)&=w_{0}+\frac{1}{2}w_{2}r^2+\cdots,\nonumber\\
	X(r)&=x_{0}+\frac{1}{2}x_{2}r^2+\cdots.
\end{align} 
If we substitute Eq.~\eqref{eq:zero} into Eqs.~\eqref{eq:H1}, \eqref{eq:K}, \eqref{eq:W} and \eqref{eq:X} and solve the equations order by order, we can obtain the first order constraints $\mathcal{O}(r^0)$ and the second order constraints $\mathcal{O}(r^2)$. In particular,  the first order constraints $\mathcal{O}(r^0)$, following these relations can be written as
\begin{align}
	x_0&=(\rho_0+p_0)e^{\nu_0/2}\biggl\{\bigg[\frac{4\pi}{3}(\rho_0+3p_0)\nonumber\\
	&\quad-\omega^2e^{-\nu_0}/\ell\bigg]w_0+\frac{1}{2}k_0\biggr\}\,,\label{eq:regular11}\\
	y_0&=\frac{2\ell k_0+16\pi(\rho_0+p_0)w_0}{\ell(\ell+1)}\,.\label{eq:regular12}
\end{align}
where $\rho_{0}$, $p_{0}$ and $\nu_{0}$ are constants defined in the power series as
\begin{align}
	\rho&=\rho_{0}+\frac{1}{2}\rho_{2}r^2+\cdots\,,\\
	p&=p_{0}+\frac{1}{2}p_{2}r^2+\frac{1}{4}p_{4}r^4+\cdots,\\
	\nu&=\nu_{0}+\frac{1}{2}\nu_{2}r^2+\frac{1}{4}\nu_{4}r^4+\cdots.
\end{align}
The constants $p_2$, $p_4$, etc., are obtained from the series expansion of Eqs.~\eqref{eq:tov1}\eqref{eq:tov2}\eqref{eq:tov3}. They are 
\begin{align}
	p_2=&-\frac{4\pi}{3}(\rho_0+p_0)(\rho_0+3p_0)\,,\\
	\rho_2=&\frac{p_2(\rho_0+p_0)}{\gamma_0 p_0}\,,\\
	\nu_2=&\frac{8\pi}{3}(\rho_0+3p_0)\,,\\
	p_4=&-\frac{2\pi}{5}(\rho_0+p_0)(\rho_2+5p_2)-\frac{2\pi}{3}(\rho_2+p_2)(\rho_0+3p_0)\nonumber\\
	&-\frac{32\pi^2}{9}\rho_0(\rho_0+p_0)(\rho_0+3p_0)\,,
  \end{align}
  and
\begin{align}
	\nu_4=&\frac{4\pi}{5}(\rho_2+5p_2)+\frac{64\pi^2}{9}\rho_0(\rho_0+3p_0)\,.
\end{align}
The second order constraints $\mathcal{O}(r^2)$ imposed on these functions $h_2$, $k_2$, $w_2$ and $x_2$ are
\begin{widetext}
	\begin{align}
	-\frac{1}{4}(\rho_0+p_0)k_2&+\frac{1}{2}\left[p_2+(\rho_0+p_0)\frac{\omega^2(\ell+3)}{\ell(\ell+1)}e^{-\nu_0}\right]w_2+\frac{1}{2}e^{-\nu_0/2}x_2=\frac{1}{4}\nu_2e^{-\nu_0/2}x_0+\frac{1}{4}(\rho_2+p_2)k_0+\frac{1}{4}(\rho_0+p_0)Q_0\nonumber\\
	&+\frac{1}{2}\omega^2(\rho_0+p_0)e^{-\nu_0}Q_1-\left\{p_4-\frac{4\pi}{3}\rho_0p_2+\frac{\omega^2}{2\ell}[\rho_2+p_2-(\rho_0+p_0)\nu_2]e^{-\nu_0}\right\}w_0\,,\\
	\frac{1}{2}(\ell+2)k_2&-\frac{1}{4}\ell(\ell+1)y_2+4\pi(\rho_0+p_0)w_2=\frac{4\pi}{3}(\rho_0+3p_0)k_0+\frac{1}{2}Q_0-4\pi\left[\rho_2+p_2+\frac{8\pi}{3}\rho_0(\rho_0+p_0)\right]w_0\,,\\
	\frac{1}{2}(\ell+3)y_2&-k_2-8\pi(\rho_0+p_0)\frac{\ell+3}{\ell(\ell+1)}w_2=4\pi\left[\frac{1}{3}(2\ell+3)\rho_0-p_0\right]y_0+\frac{8\pi}{\ell}(\rho_2+p_2)w_0-8\pi(\rho_0+p_0)Q_1+\frac{1}{2}Q_0\,,\\
	\frac{1}{2}(\ell+2)x_2&-\frac{1}{8}\ell(\ell+1)(\rho_0+p_0)e^{\nu_0/2}y_2-(\rho_0+p_0)e^{-\nu_0/2}\left[\frac{1}{4}(\ell+2)\nu_2-2\pi(\rho_0+p_0)-\frac{1}{2}\omega^2e^{-\nu_0}\right]w_2\nonumber\\
	=&\frac{1}{2}\left[\rho_2+p_2+\frac{1}{2}(\rho_0+p_0)\nu_2\right]\frac{\ell x_0}{\rho_0+p_0}+(\rho_0+p_0)e^{\nu_0/2}\biggl\{\frac{1}{2}\nu_2k_0+\frac{1}{4}Q_0+\frac{1}{2}\omega^2e^{-\nu_0}y_0-\frac{1}{4}\ell(\ell+1)\nu_2Q_1\nonumber\\
	&+\left[\frac{1}{2}(\ell+1)\nu_4-2\pi(\rho_2+p_2)-\frac{16\pi^2}{3}\rho_0(\rho_0+p_0)+\frac{1}{2}\left(\nu_4-\frac{4\pi}{3}\rho_0\nu_2\right)+\frac{1}{2}\omega^2e^{-\nu_0}\left(\nu_2-\frac{8\pi}{3}\rho_{0}\right)\right]w_0\biggr\}\,,
	\end{align}
\end{widetext}
where 
\begin{align}
	Q_{0}&=\frac{4}{(\ell+2)(\ell-1)}\biggl\{8\pi e^{-\nu_0/2}x_0-\left(\frac{8\pi}{3}\rho_{0}+\omega^2 e^{-\nu_0}\right)k_0\nonumber\\
	&\quad-\left[\frac{2\pi}{3}\ell(\ell+1)(\rho_{0}+3 p_{0})-\omega^2 e^{-\nu_0}\right]y_0\biggr\}\,,
\end{align}
\begin{align}
	Q_{1}&=\frac{2}{\ell(\ell+1)}\left[\frac{x_0}{\gamma_0p_0}e^{-\nu_0/2}+\frac{3}{2}k_0+\frac{4\pi}{3}(\ell+1)\rho_0w_0\right]\,.
\end{align}
In summary, for the even-parity modes, we have four  equations Eq.~\eqref{eq:H1}, \eqref{eq:K}, \eqref{eq:W} and \eqref{eq:X} governing the ``evolution" of four physically independent perturbative variables $H_1,K,W,X$ in the radial direction. To obtain physically relevant solutions, we also need to specify relevant boundary conditions. For example, $X(R)=0$ must be satisfied because the pressure must vanish on the surfaces. In addition,  we have two regular boundary conditions Eq.~\eqref{eq:regular11} and Eq.~\eqref{eq:regular12} to be satisfied at the center of the star. Therefore there are in total three boundary conditions at the star center and the star surface. As a result, the four free variables  can be constrained  leaving with one free parameter, which should be determined by using the information of perturbations outside the star.
\subsubsection{Odd parity}

In the interior region of a star, the odd parity modes are described by a single wave equation for gravitational perturbations. The odd parity modes can not couple to the pulsation of the star \cite{thorne1967non}.
The master wave variable $X^{\rm int}$ is related to the functions $h_{1}$ and $h_{0}$ by \cite{nakamura1987general}
\begin{align}
    h_{1} &= e^{\lambda}rX^{\rm int},\\
    h_{0} & = \frac{i}{\omega}e^{-\lambda}\frac{d}{dr}(rX^{\rm int})\,.
\end{align}
The equation governing the wave function $X$ is derived from the linearized Einstein equation for odd parity.
\begin{align}
    \frac{d^2X^{\rm int}}{dr^{*2}}+(\omega^2-V^{\rm int})X^{\rm int}=0,
\end{align}
where
\begin{align}
    r^{*}&=\int_{0}^{r}e^{-(\nu-\lambda)/2}dr \,,\nonumber\\
    V^{\rm int}&=e^{\nu}\left[\frac{\ell(\ell+1)}{r^2}-\frac{6M(r)}{r^3}-4\pi(p-\rho)\right] \,.\nonumber
\end{align}
\subsection{Perturbations outside the star}\label{sec:pout}
Outside the star, only gravitational perturbations are relevant, which are described by a second-order wave equation and a master variable $Z_{\rm even, \rm odd}$ for the even and odd parity case respectively 
\begin{align}\label{eq:perturbout}
	\left[\frac{d^2}{dr*^2}+\omega^2-V_{\rm even, \rm odd}\right]Z_{\rm even, \rm odd}=S_{\rm even, \rm odd}\,,
\end{align}
where $V$ is an effective potential defined as
\begin{align}
	V_{\rm even}&=\frac{e^{-\lambda}[2n^2(n+1)r^3+6n^2Mr^2+18nM^2r+18M^3]}{r^3(nr+3M)^2}\,,\\
        V_{\rm odd}&=e^{\lambda}\left(\frac{\ell(\ell+1)}{r^2}-\frac{6M}{r^3}\right)\,.
\end{align}
Here $S$ is the  source term for the wave equations. In this work, they are obtained by using the test particle trajectory orbiting around a massive star. The detailed expression is given in Appendix. \ref{sec:st}.

The master variable $Z$ (for even and odd parity respectively, subscript abridged) must match with internal metric perturbation quantities at the stat surface $r=R$. For the even parity, the Zerilli function $Z$ is related to interior perturbation functions at surface as
\begin{align}
	Z(r^*)&=-\frac{r^2e^{-\lambda}}{nr+3M}H_1+\frac{r^2}{nr+3M}K\,,\label{eq:boundary21}\\
	\frac{dZ(r^*)}{dr^*}&=\frac{n(n+1)r^2+3nMr+6M^2}{(nr+3M)^2}e^{-\lambda}H_{1}\nonumber\\
 &\quad-\frac{nr^2-3nMr-3M^2}{(nr+3M)^2}K\,,\label{eq:boundary22}
\end{align}
which is evaluated at $r=R$.

For the odd parity, metric perturbation $h_{1}$ and $h_{0}$ is related to the single wave variable $Z$ through
\begin{align}
    h_{1} &= e^{\lambda}rZ,\\
    h_{0} & = \frac{i}{\omega}e^{-\lambda}\frac{d}{dr}(rZ)\nonumber\\
    &\quad-\frac{8\pi i}{\omega}\frac{r^2 e^{-\lambda}}{\sqrt{2n(n+1)}}D(\omega, r),
\end{align}
where $D(\omega, r)$ is one of the source terms, as detailed in Appendix. \ref{sec:st}. This expression should also be evaluated at $r=R$.

Away from the star surface, in order to solve Eq.~\eqref{eq:perturbout} with boundary conditions \eqref{eq:boundary21} and \eqref{eq:boundary22}, we apply the Green's function approach. We first consider the homogeneous of Eq.~\eqref{eq:perturbout}
\begin{align}\label{eq:homogeneous}
\left[\frac{d^2}{d{r^{*}}^2}+\omega^2-V\right]Z^{\rm hom}=0\,,
\end{align}
with the same boundary conditions. There are two independent solutions $Z^{\rm hom}_{\rm out}$ and $Z^{\rm hom}_{\rm in}$ which correspond to outgoing and incoming waves at infinity respectively,
\begin{align}
    &Z^{\rm hom}_{\rm out}\rightarrow e^{i\omega r^*},\\
    &Z^{\rm hom}_{\rm in}\rightarrow e^{-i\omega r^*},
\end{align}
as $r \rightarrow \infty$. Therefore, the exact solution of Eq.~\eqref{eq:homogeneous} can be expressed as
\begin{align}
    Z^{\rm hom}=\alpha Z^{\rm hom}_{\rm out}+\beta Z^{\rm hom}_{\rm in},
\end{align}
where $\alpha$ and $\beta$ should be compatible with the boundary conditions \eqref{eq:boundary21} and \eqref{eq:boundary22}.

With the homogeneous solutions and the boundary conditions \eqref{eq:boundary21} and \eqref{eq:boundary22},  the solution of the wave equation \eqref{eq:perturbout} is given by \cite{arfken2011mathematical} 
\begin{align}
    Z(r^*)&=\alpha Z^{\rm hom}_{\rm out}(r^*)+\beta Z^{\rm hom}_{\rm in}(r^*)\nonumber\\
    &\quad+\int^{\infty}_{R^*}G(r^*,s^*)S(s^*)ds^*,
\end{align}
where $G(r^*,s^*)$ is Green's function: 
\begin{align}
    G(r^*,s^*)&=\frac{1}{W}\bigg[-Z^{\rm hom}_{\rm out}(r^*)Z^{\rm hom}_{\rm in}(s^*)\nonumber\\
    &+Z^{\rm hom}_{\rm in}(r^*)Z^{\rm hom}_{\rm out}(s^*)\bigg]\Theta(r^*-s^*)\,.
\end{align}
Here $W$ is the Wronskian and $\Theta(x)$ is the Heaviside function. This Green's function corresponds to the inhomogeneous equation \eqref{eq:perturbout} with a $\delta$ function source and the boundary conditions:
\begin{align}
    Z(R)&=0\,,\\
    \frac{dZ}{dr^*}(R)&=0\,.
\end{align}
At spatial infinity, the solution $Z(r^*)$ is 
\begin{align}
    Z(r^*\rightarrow\infty)=(\alpha+\tau)e^{i\omega r^*}+(\beta+\sigma)e^{-i\omega r^*}\,,
\end{align}
where 
\begin{align}
    \tau&=-\frac{1}{W}\int_{R}^{\infty}Z^{\rm hom}_{\rm in}(s^*)S(s^*)ds^*\,,\\
    \sigma&=\frac{1}{W}\int_{R}^{\infty}Z^{\rm hom}_{\rm out}(s^*)S(s^*)ds^*\,.
\end{align}
Since the wave at infinity must be  outgoing, we have
\begin{align}
    \beta+\sigma=0.
\end{align}
Therefore at infinity, the asymptotic behaviour of $Z(r^*)$ is given by
\begin{align}
    Z(r^*)\rightarrow A e^{i \omega r^*}, \quad {\rm for}\  r\rightarrow\infty,
\end{align}
where the amplitude $A_{lm}$ is
\begin{align}
    A&=\alpha+\tau=-\frac{1}{\beta}\left(-\beta\tau+\alpha\sigma\right)\,\nonumber\\
    &=-\frac{1}{\beta}\frac{1}{W}\int_{R}^{\infty}\left(\alpha u^{\rm hom}_{\rm out}+\beta u^{\rm hom}_{\rm in}\right)S(s^*)ds^*\,\nonumber\\
    &=-\frac{1}{\beta}\frac{1}{W}\int^{\infty}_{R^*}Z^{\rm hom}(s^*)S(s^*)ds^* \nonumber \\
    &= -\frac{1}{W}\int^{\infty}_{R^*}\left (\frac{\alpha}{\beta} Z^{\rm hom}_{\rm out}+ Z^{\rm hom}_{\rm in}\right )S(s^*)ds^*\,.
\end{align}
Notice that the ratio $\alpha/\beta$ is fully determined by requiring $Z^{\rm hom}$ to satisfy the boundary conditions \eqref{eq:boundary21} and \eqref{eq:boundary22}.

\section{Gravitational self force}
In order to determine the long-term evolution of a particle moving around a central body, it is necessary to understand the radiation reaction, i.e., the gravitational self-force. 
The force includes both the conservative part and the dissipative part, where the conservative part is relevant for the self-force modified conserved quantities, e.g. the energy, and the dissipative part being is related to the gravitational wave flux.  In this work, we only consider circular orbits, so that the dissipative effect only affects $F^{t}, F^{\varphi}$, and the conservative part is characterized by $F^{r}$ \cite{barack2007gravitational}. To address the finite size effect of the central body, we use various star models to evaluate the self force. One interesting observation we have made is that there are universal relations for the self force quantities, which to the leading order only depends on the dynamical tidal deformability. This is convenient as we can use only two numbers to summarize the finite-size effect: the equilirium tidal Love deformability and the f-mode frequency. We start with tide-induced gravitational wave flux for the $\ell=2$, including both $m=\pm2$ modes, and then discuss gravitational self force $F^{t}$ and $F^{r}$ for $\ell=2$ perturbations, including the demonstration of universal relations. In Sec.~\ref{sec:ho} we also  discuss higher order contributions with $\ell\geq3$ to the gravitational self force.
\subsection{Tide-induced gravitational wave flux}

In this section, we consider the tide-induced gravitational wave flux $P^{\rm tide}$. We are interested in computing the difference between the energy flux generated by a point mass circulating around a compact star and the one generated by a point mass circulating around a Schwarzschild black hole (the masses are the same) at infinity with the same orbital frequency, which is defined as the tidal induced gravitational wave flux. 

As the central compact star can have various size, internal structure and equation of state, it is important to efficiently characterize the tidal induced flux for various compact stars, especially for the waveform model construction purpose. Intuitively the deformed star produces additional radiative moments in the radiative zone, which superposes on the radiaitve moments due to the moving point mass and affects the total flux. It is also reasonable to expect that the star deformation and the resulting radiative moments is related to the dynamical tidal deformability of the star. 
At the quadrupole order, the star deformation (in the static limit) can be described as
\begin{align}
    Q_{ij}=-\tilde{\lambda}_{2}\mathcal{E}_{ij}\,,
\end{align}
where $Q_{ij}$ is the quadrupole moment of the star, and $\mathcal{E}_{ij}$ is the external quadrupole tidal field. The $\ell=2$ static dimensionless Love number is given by
\begin{align}
    k_{2}=\frac{3}{2}\tilde{\lambda}_2 R^{-5} = \frac{3}{2}\lambda_2 \left (\frac{M}{R} \right )^{5}\,,
\end{align}
where $\lambda_2 =\tilde{\lambda}_2 M^{-5}$ is the dimensionless tidal deformability.

\begin{figure*} 
    \centering
    \includegraphics[height=12cm,width=17.5cm]{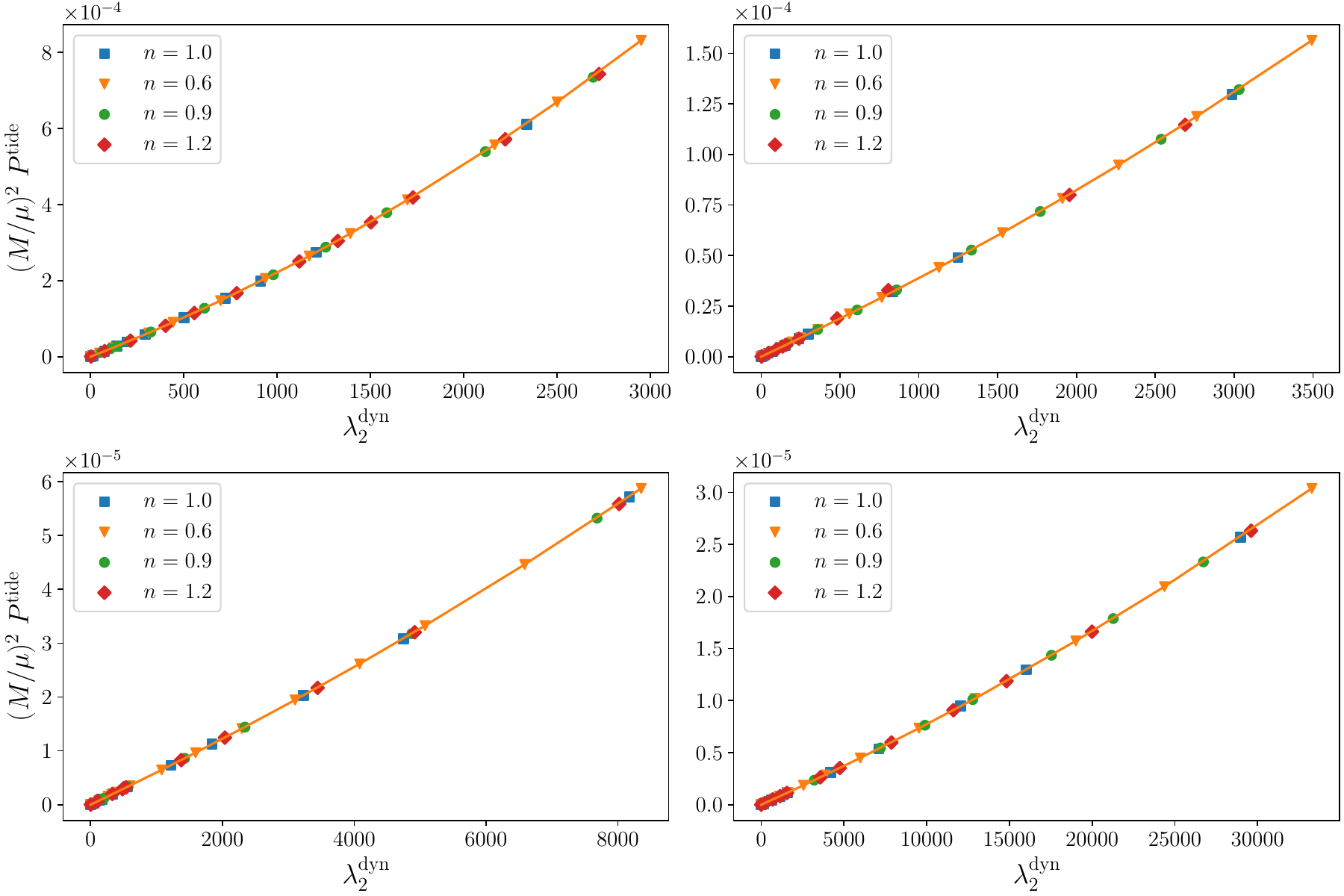}
    \caption{The tide-induced gravitational wave flux for $l=2$, including both $m=\pm 2$ modes, denoted as $P^{\rm tide}$
 , are illustrated for various dynamical Love numbers across four distinct panels. Each panel corresponds to a specific orbital frequency $M\Omega$, namely 0.0516, 0.0413, 0.032, and 0.0237. The range of dynamical deformability for the plots is chosen to show certain level of nonlinearity. The results indicate that for a given orbital frequency, the tide-induced gravitational wave flux can be universaly characterized by the (dynamical) tidal Love number of the compact star. For these plot we have chosen EOS with four different polytropic indexes, $n=1.0$, $n=0.6$, $n=0.9$ and $n=1.2$. These  stars have the same mass $1.4M_{\odot}$.}
    \label{fig:flux1}
\end{figure*}

Therefore for weak perturbations of any compact star we first propose an ansatz for the tide-induced flux as 
\begin{align}\label{eq:Ptide}
    P^{\rm tide}(\omega)=\lambda_2^{\rm dyn}P_{0}(\omega),
\end{align}
where $P_{0}(\omega)$ is a function of the frequency $\omega=m\Omega$. On the other hand, the 
dynamic (dimensionless) tidal deformability is given by \cite{bini2012effective,chakrabarti2013effective}
\begin{align}\label{eq:dyn_lambda}
    \lambda_2^{\rm dyn}=\lambda_2\frac{\omega_{f}^2}{\omega_{f}^2-\omega^2}\,,
\end{align}
where $\lambda_2=\frac{2}{3} k_{2}R^{5}/M^5$ is the dimensionless equilibrium tidal deformability and $\omega_{f}$ is the f-mode frequency of the compact star. By examining the flux for different star configurations, we  find that the finite size effect of the star are nicely encoded in $\lambda_2^{\rm dyn}$, leaving an approximately universal $P_0(\omega)$ for various stars. This property is very important for constructing an efficient waveform model for various kinds of black hole mimickers, as they may be quite different in nature. Another interesting fact we  find is that the linear dependence in $\lambda_2^{\rm dyn}$ may not be fully accurate (i.e., see Fig.~\ref{fig:flux1}) when $\omega$  increases relative to intrinsic frequencies of the star, possibly due to the break-down of the single-mode approximation in characterising the star deformation. Nevertheless we find a power-law expansion in $\lambda_2^{\rm dyn}$ gives rise to a decent universal description as well.

In order to determine the f-mode frequency of a neutron star, we apply the approximate relationship between the f-mode frequency and the tidal deformability as provided in \cite{chan2014multipolar,shashank2023f}. The explicit relationship is given by
\begin{align}\label{eq:fmode}
    M\omega_{f}=\Sigma_{i}a_{i}(\lambda_2)^{i},
\end{align}
where 
\begin{align}
    a_0&=1.442\times 10^{-1},\quad a_1=3.005\times 10^{-2}, \nonumber\\
    a_2&=-1.607\times 10^{-2},\ a_3=2.092\times 10^{-3},  \nonumber\\
    a_4&=-9.247\times 10^{-5}. \nonumber
\end{align}
It is worth to note that the f-mode frequency calculated using Eq.\eqref{eq:fmode} typically has an error of approximately $1\%$ \cite{chan2014multipolar,shashank2023f}. This error becomes significantly amplified when it is applied in Eq.\eqref{eq:dyn_lambda} near resonance. As a result, we numerically refine the values obtained from Eq.~\eqref{eq:fmode} by numerically searching for the resonant frequency, using Eq.\eqref{eq:dyn_lambda} to compute the initial values of the searching algorithm.

In order to test the universal relations between tidal-induced flux and the dynamic tidal Love number, we need to explore different star EOS. For simplicity we pick the  polytropic equation of state:
\begin{align}
    p=\kappa\rho^{1+1/n},
\end{align}
where $\kappa$ is the polytropic constant and $n$ is the polytropic index. In this work we have chosen four different polytropic indexes, $n=1.0$, $n=0.6$, $n=0.9$ and $n=1.2$. For each $n$, we vary the values of $\kappa$ and central density $\rho_{c}$ of neutron star to obtain different values of $\lambda_2^{\rm dyn}$.

\begin{figure*}
    \centering
    \includegraphics[height=6cm,width=17cm]{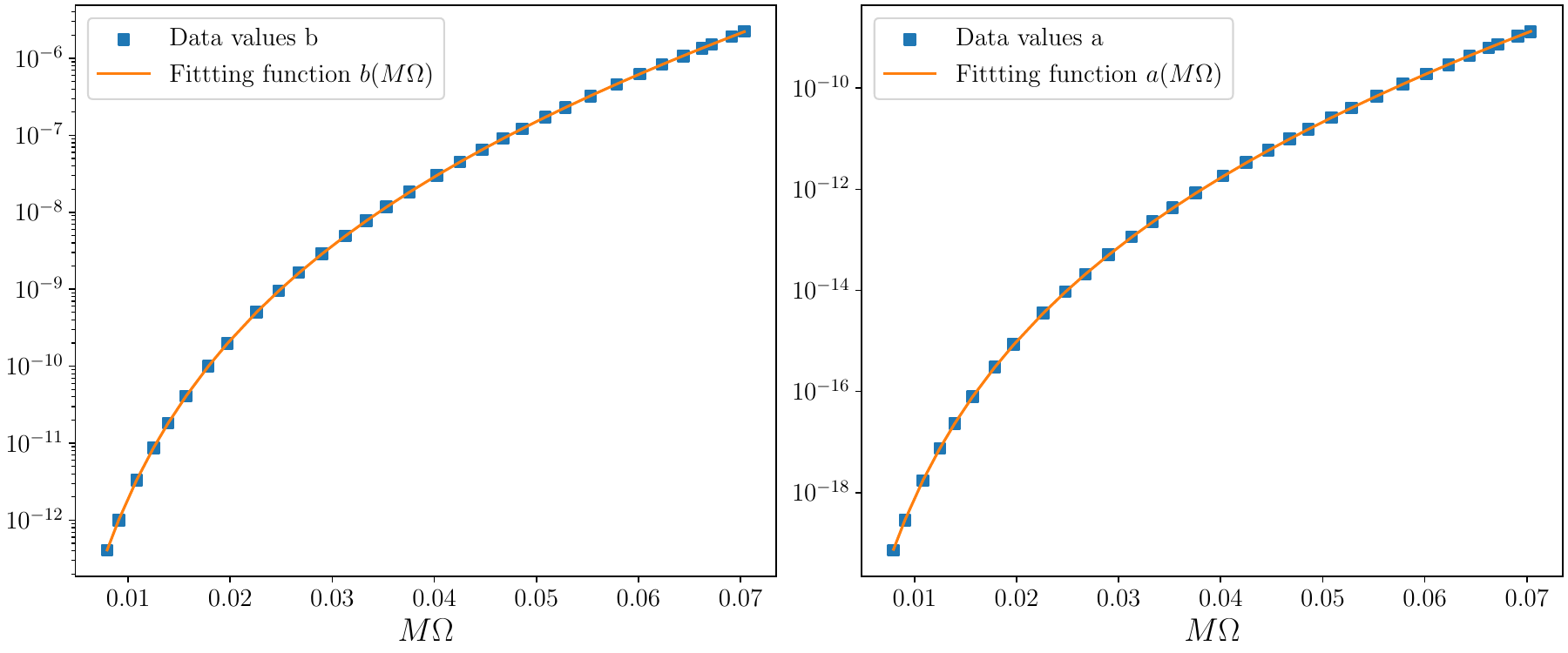}
    \caption{Data values $a$, $b$ and the corresponding fitting functions $a(M\Omega)$, $b(M\Omega)$ are shown. We use several  stars with polytropic equations of state $n=0.6$. For different values of angular frequency $\Omega$, the associated values of $a$, $b$ can be similarly obtained.}
    \label{fig:ab}
\end{figure*}

\begin{figure} 
    \centering
    \includegraphics[height=6.3cm,width=9.2cm]{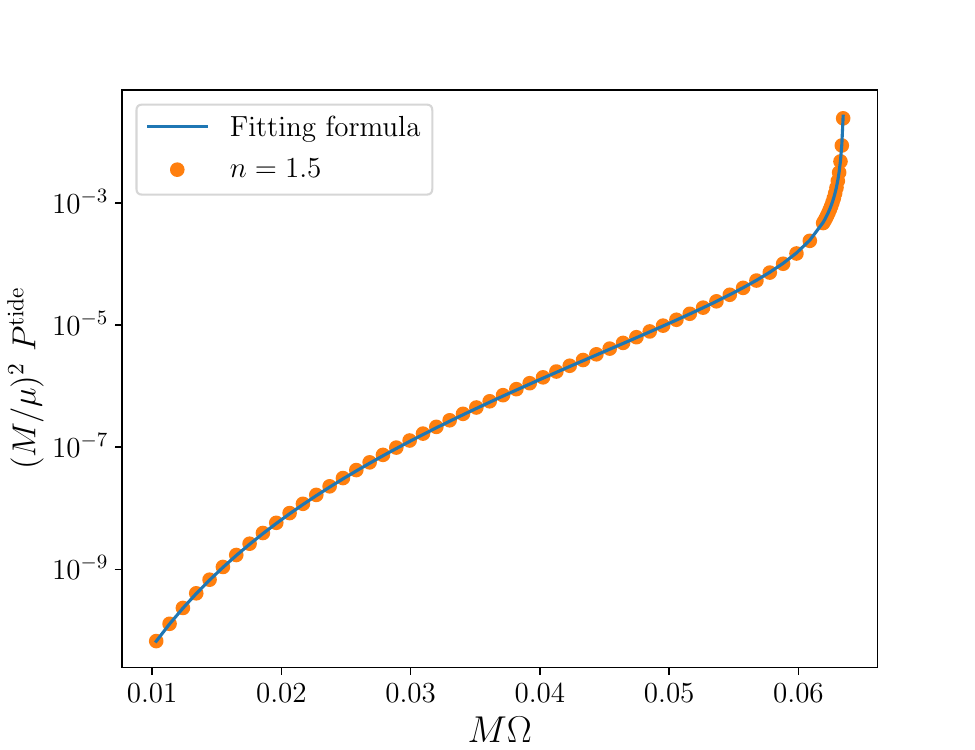}
    \caption{Comparison between the fitting formula Eq.~\eqref{eq:fitting} of tide-induced energy flux $P^{\rm tide}$ at infinity and that produced by a point particle orbiting around a compact star with polytropic equation of state $n=1.5$, $\rho_{c}=5\times10^{15}\,{\rm g/cm^3}$ and $\kappa=12.445425\,{\rm km^2}$. In the most range of inspiral process, the error is less than $3\%$. }
    \label{fig:comparison}
\end{figure}

Now we are ready to explore the tide-induced flux for various types of central stars and at different orbit frequencies of the point mass. We first fix the orbital frequency $\Omega$, and compute the tide-induced gravitational wave flux $P^{\rm tide}$ $l=2$, including both $m=\pm 2$ modes at infinity as shown in Fig.~\ref{fig:flux1}. We pick different polytropic parameters $n,\kappa$ but vary the central density to make sure star mass remains constant. For all the frequencies we have explored, the tide-induced gravitational flux $P^{\rm tide}$ varies on a single curve as a function of the dynamic tide deformability. This is nontrivial as different points lying on the curve may belong to different star EOS, as shown explicitly in Fig.~\ref{fig:flux1} for four representative frequencies. The curve may deviate from linearity for large $\lambda_2^{\rm dyn}$. We suspect that this nonlinear behavior may come from the additional contribution from higher-order p-modes which generally have higher resonant frequency that the f-mode. As the ratio between the orbital frequency $\Omega$ and the f-mode frequency $\omega_f$ increases, the  higher order modes may also be increasingly excited. Therefore for the same $\Omega$ but larger $\lambda_2^{\rm dyn}$ (so that the star size is larger and $\omega_f$ is smaller), the nonlinear effect becomes larger.
On the other hand, in a recent work by Pitre and Poisson \cite{pitre2024general}, the time-domain tidal response of a star is compared to a mode decomposition picture. 
Their results show that the  dynamical tides approximately matches that of the f-mode description in the mode representation. The relative error is about $5\%$.

With the observation of universality shown in Fig.~\ref{fig:flux1}, it is natural to expect that tide-induced gravitational wave flux is be well characterized by the dynamic tidal deformability of the compact star and orbital frequency. Therefore the next step is to  obtain a mathematical description for the universal relation between tide-induced gravitational wave flux and tidal deformability, regardless of equation of state. Based on the discussion of nonlinearity, we further assume the following fitting formula 
\begin{align}\label{eq:fitting}
    P^{\rm tide}=(\lambda_2^{\rm dyn})^2a(M\Omega)+\lambda_2^{\rm dyn} b(M\Omega)\,,
\end{align}
with the functions $a(M\Omega), b(M\Omega)$ to be determined by the numerical data with different $\lambda_2^{\rm dyn}$ and $\Omega$.

For simplicity we choose a $n=0.6$ type of star with various central density and fixed orbital frequency, similar to Fig.~\ref{fig:flux1}. After applying to linear $+$ quadratic fit for $\lambda_2^{\rm dyn}$, we can obtain $a(M\Omega), b(M\Omega)$ for that particular frequency. After that we vary the orbital frequency to compute frequency-dependent data points for $a(M\Omega)$ and $b(M\Omega)$, which are shown in Fig.~\ref{fig:ab}. We can fit these data points using rather simple fitting formulas as
\begin{align}\label{eq:fitting2}
    a(M\Omega)=&\frac{32}{5}(M\Omega)^{30/3}\left(3.648\, e^{40.48(M\Omega)}+ 6.737\right),\\
    b(M\Omega)=&\frac{32}{5}(M\Omega)^{20/3}\left(1.405\, e^{30.63(M\Omega)}+4.614\right).
\end{align}

In order to assess the accuracy of these fitting formulas, we pick a star with a different EOS, e.g. $n=1.5$, that has not been previously considered in this study. The detailed parameter of the star and the tide-induced energy flux produced by an orbiting point mass with various frequencies are shown in Fig.~\ref{fig:comparison}. These data points are also compared with the fitting formula  in Eq.~\eqref{eq:fitting2}. It's evident that throughout most of the inspiral process, the relative error is bounded below $3\%$, even up to an angular frequency of $M\omega = 0.1267$  that is close to the f-mode frequency $(M\omega_{f}=0.1271)$. We have tested several star configurations with various polytropic EOS and the performance of the fitting formula is similar. In the future it may be worth to further test this with more general star EOS, and different kinds of central object such as boson stars.

There is another intriguing observation associated with the mathematical form of Eq.~\eqref{eq:fitting}, that we have not included a term that does not depend on $\lambda_2^{\rm dyn}$. The physical meaning of this term would be the tide-induced flux in the limit that the dynamical tidal deformability approaches zero.
Numerically in such limit the data points suggest that the tide-induced flux actually decreases to zero within the numerical precision of our implementation. Consequently, in this extreme scenario, the energy flux radiated to infinity is the same for the case with a point particle  orbiting around a compact star (with $\lambda_2^{\rm dyn}$ approaching zero) and that with a point particle orbiting a black hole. This point is rather surprising as intuitively the compact star does not have a horizon, so the ingoing gravitational wave should eventually come out and superpose on the original outgoing gravitational waves, modifying the total outgoing flux. Indeed this point should be further tested in the future with higher numerical precision.

This finding is actually consistent the result presented in \cite{datta2020tidal}, where a Teukolsky equation is solved with modified boundary condition near the horizon. It is demonstrated that the energy flux at infinity remains unchanged irrespective of the value of $\mathcal{R}$, which includes the black hole case $(\mathcal{R}=0)$. Here 
$\mathcal{R}$ denotes the {\it ad hoc} reflection coefficient defined on the inner surface.
\subsection{Gravitational self force for $\ell=2$ perturbations}

We have shown that the tide-induced modification of the gravitational wave flux radiated to infinity satisfies an universal relation with the dynamic (dimensionless) tidal deformability of the central object. In this section, we switch the focus from the field at infinity to the field near the orbiting point mass, although they are related by the wave equation, and show that similar universal relation also holds for the tide-induced gravitational self force. We only consider the $\ell=2$ piece of the perturbation here because it is the dominant part of the flux, and that we only use $k_2$ to describe the dynamic tide.


Gravitational self force is the necessary ingredient to build the waveform of EMRIs. Before  showing the calculation details,
We first briefly review the basic strategy of evaluating  the gravitational self force for a point particle $\mu$ in a circular orbit around Schwarzschild black hole $M$, with $\mu/M \ll 1$. The spacetime metric with a moving point mass can be written as 
\begin{align}
    g_{ab}=g^{0}_{ab}+h_{ab}\,,
\end{align}
where $g^{0}_{ab}$ is the Schwarzschild metric, and $h_{ab}$ is a metric perturbation.
Based on the formalism in  \cite{pound2014gravitational}, we can write down the general gravitational self force as
\begin{align}
    F^{a}=-\frac{1}{2}\mu (g_{0}^{ab}+\Tilde{u}^{a}\Tilde{u}^{b})(2\nabla_{d}h_{bc}-\nabla_{b}h_{cd})\Tilde{u}^{c}\Tilde{u}^{d}\,,
\end{align}
where $\Tilde{u}^{a}$ is a smooth extension of the four-velocity off of the particle’s worldline. As  the gravitational self force formally diverges near the point mass,  the metric perturbation into a singular piece and a regular piece\cite{detweiler2003self}
\begin{align}
    h_{ab}=h_{ab}^{\rm S}+h_{ab}^{\rm R}\,,
\end{align}
where 
\begin{align}
    -2\delta G_{ab}[h^{\rm S}]&=-16\pi T_{ab}\,,\nonumber\\
    -2\delta G_{ab}[h^{\rm R}]&=0\,.\nonumber
\end{align}
In the above equation, $h_{ab}^{\rm S}$ is singular at the location of particle, but does not contribute to the gravitational self force. On the other hand $h_{ab}^{\rm R}$ is regular on the world line and can be used to compute the self force:
\begin{align}
    F_{\rm self}^{a}=-\frac{1}{2}\mu (g_{0}^{ab}+{u}^{a}{u}^{b})(2\nabla_{d}h_{bc}^{R}-\nabla_{b}h_{cd}^{R}){u}^{c}{u}^{d}\,.
\end{align}
The equation of motion of a particle $\mu$ is given by
\begin{align}\label{eq:equa_motion}
    \mu\frac{D^2x^{a}}{D\tau^2}=\mu\frac{Du^{a}}{D\tau}=F^{a}\,,
\end{align}
where the covariant derivatives are taken with respect to the background Schwarzschild spacetime. Then we define $\mathcal{E}$ and $\mathcal{L}$ as follows:
\begin{align}
    \mathcal{E}=-u_{t}\,,\quad\,\mathcal{L}=u_{\varphi}\,.
\end{align}
If we only consider circular orbit, by combining Eq.~\eqref{eq:equa_motion} and the normalized condition $u_{a}u^{a}=-1$, we have
through $O(\mu)$ \cite{barack2007gravitational} 
\begin{align}
    \mathcal{E}&=\mathcal{E}_{0}\left[1-\left(\frac{r_{0}}{2\mu}\right)F_{r}\right]\,,\\
    \mathcal{L}&=\mathcal{L}_{0}\left[1-\left(\frac{r_{0}^2}{2M\mu}\right)F^{r}\right]\,,
\end{align}
where $\mathcal{E}_0$ and $\mathcal{L}_{0}$ are the energy and angular momentum of circular geodesic, which are given by
\begin{align}
    \mathcal{E}_{0}&=\frac{1-2M/r_{0}}{\sqrt{1-3M/r_{0}}}\,,\\
    \mathcal{L}_{0}&=\frac{\sqrt{Mr_{0}}}{\sqrt{1-3M/r_{0}}}\,.
\end{align}
In addition, in the circular-orbit case, $F^a u_a=0$ implies that
\begin{align}
    F^{\varphi}=\frac{\mathcal{E}_{0}}{\mathcal{L}_{0}}F^{t}\,.
\end{align} 
Therefore $F^{t, \phi}$ are related, so that in this section we  focus  on the tide-induced gravitational self-force $F^{t,\rm tide}$ and $F^{r,\rm tide}$.

As mentioned above, the gravitational self force of a point particle orbiting around a Schwarzschild black hole needs to be regularized to remove the formula divergence. This often requires the computation of the singular field $h^S$. However, in our work we only consider the difference of the self force between a compact star (CS) scenario and a black hole (BH) scenario, for a point mass moving with the  {\it same} orbital frequency. As we are only interested in the leading order contribution in the mass ratio, the point mass can be assumed to follow a geodesic of the background spacetime. That means the singular field in both cases can be chosen to the {\it same}, so that
\begin{align}
h^R_{\rm CS}-h^R_{\rm BH} &= (h^R_{\rm CS}+h^S_{\rm CS})-(h^R_{\rm BH}+h^S_{\rm BH}) \nonumber \\
&= h_{\rm CS}-h_{\rm BH}\,.
\end{align}
We can just use the $h_{\rm BH}, h_{\rm CS}$ obtained from the solver described in Sec.~\ref{sec:grap} and evaluate the difference at each mode, which should be the mode decomposition of the difference in the regular field $h^R$ at the location of the  point mass.

\begin{table}[h!]
\renewcommand{\arraystretch}{1.2}
  \begin{center}
    \caption{The difference between tide-induced total energy flux and tidal induced gravitational self force $F^{t,\rm tide}$ as a function of the orbital frequency. The first column represents the orbital frequency of a point particle orbiting around a compact star or a black hole. The second and third columns show the comparison between the tide-induced self-force $F^{t,\rm tide}$ and the total tide-induced energy flux $\dot{E}_{\rm total}^{\rm tide}$ times $\Delta_{0}=-(1-2M/r_0)^{-1}u_{0}^{t}$. We  find that the fractional difference is less than $10^{-4}$ in all cases, providing a  quantitative check of our results.}
    \label{tab:comparison}
    \begin{tabular}{c | c | c} %
    \toprule[1pt]
    \makebox[0.1\textwidth][c]{$M\Omega$} & \makebox[0.18\textwidth][c]{$({M}/{\mu})^2 F^{t,\rm tide}$} & \makebox[0.18\textwidth][c]{$({M}/{\mu})^2 \Delta_{0}\dot{E}_{\rm total}^{\rm tide}$} \\
    \midrule[0.2pt]
      \hline
      0.008 & $-1.33889763\times10^{-12}$ & $-1.33888687\times10^{-12}$\\
      0.010 & $-7.68532994\times10^{-12}$ & $-7.68523929\times10^{-12}$\\
      0.015 & $-1.68805373\times10^{-10}$ & $-1.68804446\times10^{-10}$\\
      0.020 & $-1.50618037\times10^{-9}$ & $-1.50618782\times10^{-9}$\\
      0.025 & $-8.44664208\times10^{-9}$ & $-8.44672287\times10^{-9}$\\
      0.030 & $-3.57858966\times10^{-8}$ & $-3.57854450\times10^{-8}$\\
      0.035 & $-1.26088661\times10^{-7}$ & $-1.26088515\times10^{-7}$\\
      0.040 & $-3.91539506\times10^{-7}$ & $-3.91539645\times10^{-7}$\\
      0.045 & $-1.11445651\times10^{-6}$ & $-1.11445725\times10^{-6}$\\
      0.050 & $-2.99573226\times10^{-6}$ & $-2.99573072\times10^{-6}$\\
      0.055 & $-7.81094805\times10^{-6}$ & $-7.81094974\times10^{-6}$\\
      0.060 & $-2.03461381\times10^{-5}$ & $-2.03461419\times10^{-5}$\\
      0.065 & $-5.53372737\times10^{-5}$ & $-5.53372615\times10^{-5}$\\
      0.070 & $-1.73091728\times10^{-4}$ & $-1.73091727\times10^{-4}$\\
      \bottomrule[0.5pt]
    \end{tabular}
  \end{center}
\end{table}

We can now compute the gravitational wave self force as contributed by the tidal deformation of the star using the local field near the point mass. The ``t" component is dissipative, in the sense that it will flip sign with the time reversal operation.
In addition, based on the energy balance equation, the rate of change of the (specific) energy parameter $\mathcal{E}$ is
\begin{align}
    \frac{d\mathcal{E}}{dt}=-\frac{1}{\mu u_{0}^{t}}F_{t}\,.
\end{align}
It also is balanced by the flux of gravitational-wave energy radiated to infinity and through the horizon (if it exists), averaged over the orbital timescale. Then we have the following energy-balance relation ($\dot{E}_{\infty}$ and $\dot{E}_{\rm EH}$ are both positive)
\begin{align}\label{eq:Edottotal}
    \dot{E}_{\rm total}=\dot{E}_{\infty}+\dot{E}_{\rm EH}=-\mu\dot{\mathcal{E}}=\frac{F_{t}}{u_{0}^{t}}
\end{align}
for a black hole and
\begin{align}\label{eq:Edottotalcs}
  \dot{E}_{\rm total}= \dot{E}_{\infty}=\frac{F_{t}}{u_{0}^{t}}
\end{align}
for a compact star.

\begin{figure*}[htb]
\includegraphics[height=6.3cm,width=17cm]{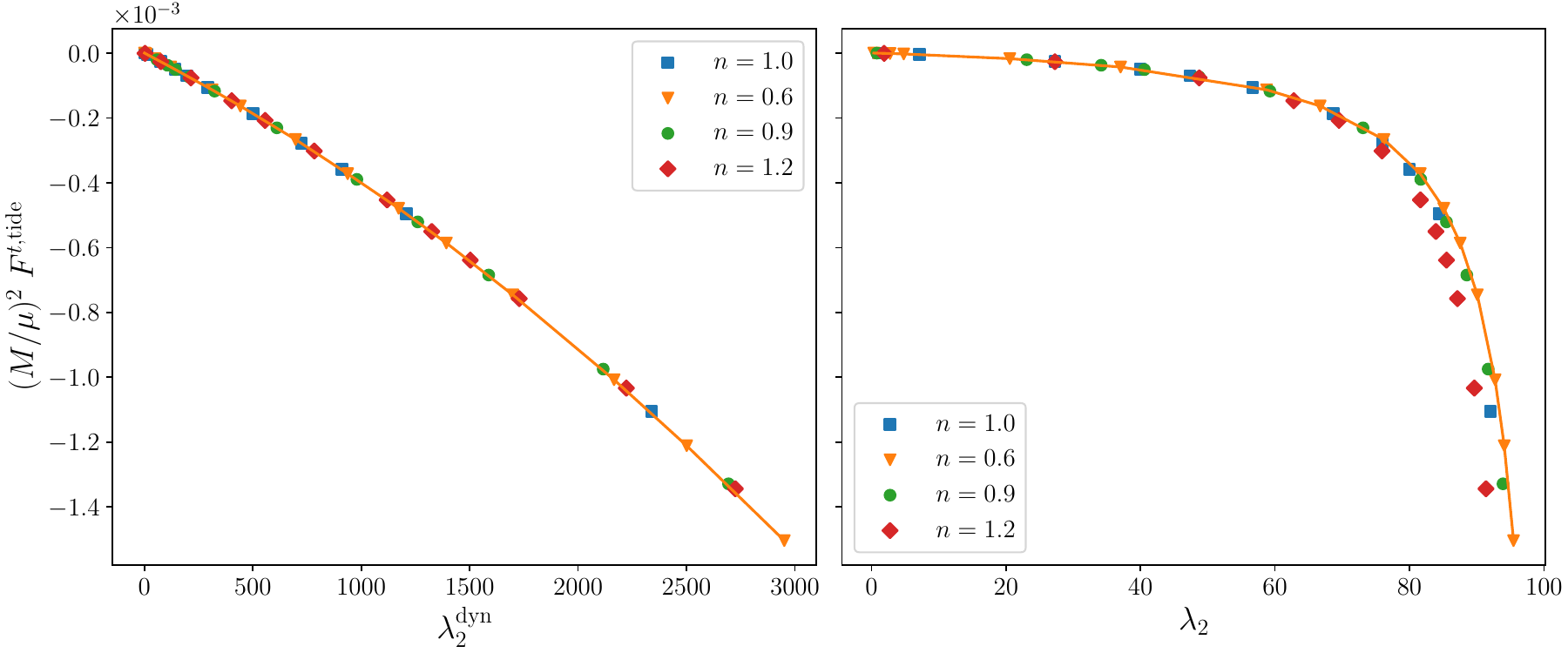}
  \caption{Tide-induced gravitational self force $F^{\rm t,tide}$ for $l=2$, including both $m=\pm 2$ modes. This is the difference between the gravitational self force produced by a point particle circular orbiting around a comapct star and a black hole with the same orbital frequency $M\Omega  = 0.0516$. We can find that $F^{\rm t,tide}$ is proportional to dynamic tidal deformability $\lambda_2^{\rm dyn}$ when $\lambda_2^{\rm dyn}$ is small, but with a small deviation towards nonlinearity can be seen as $\lambda_2^{\rm dyn}$ increase. Nevertheless, the data points from different EOS all approximately lie on the same line. On the other hand, the universal relation is broken if we switch from $\lambda_2^{\rm dyn}$ to $\lambda_2$, the equilibrium tidal deformability.}
  \label{fig:Ft}
\end{figure*}

\begin{figure*}[htb]
\includegraphics[height=6.3cm,width=17cm]{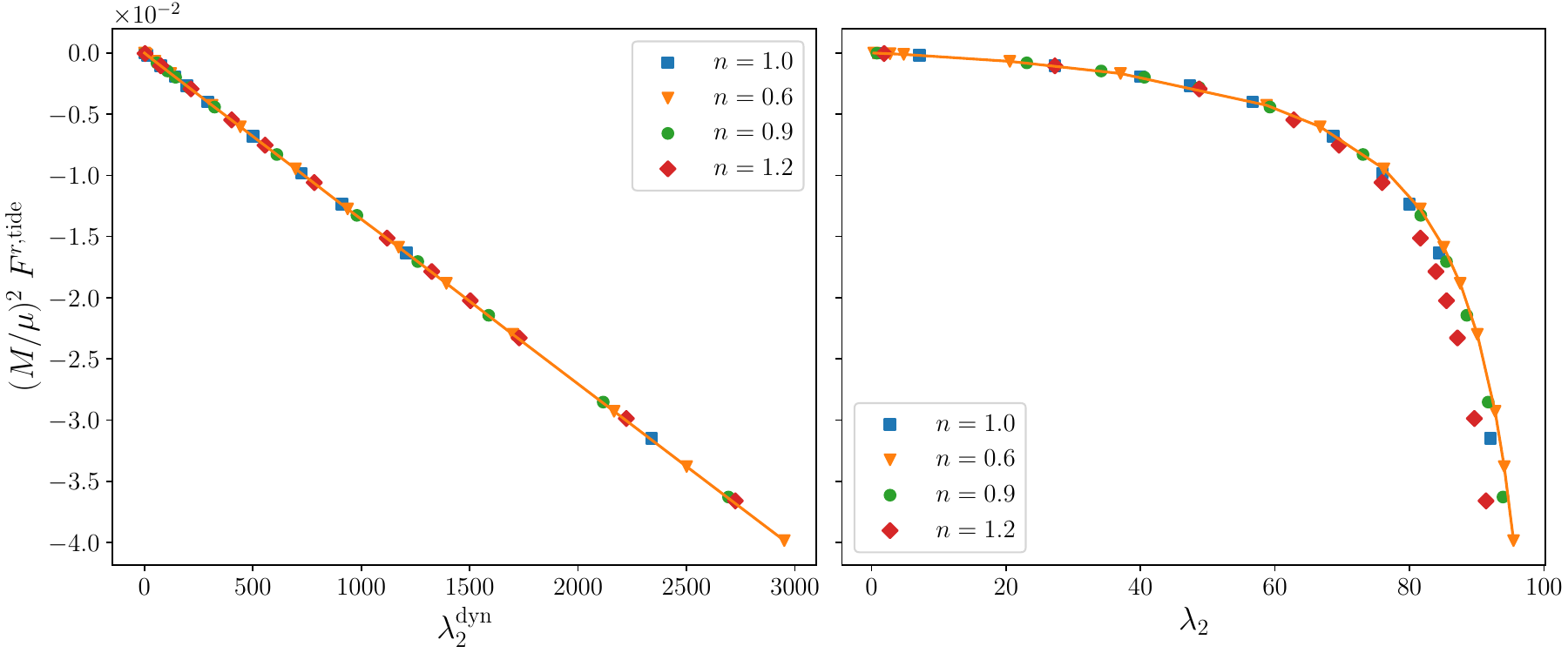}
  \caption{Tide-induced gravitational self force $F^{r, \rm tide}$ for $l=2$, including both $m=\pm 2$ modes. This is the difference between gravitational self force produced by a point particle circular orbiting around a compact star and a black hole with the same orbital frequency $M\Omega = 0.0516$. 
    For the dynamical range considered, we  find that $F^{r, \rm tide}$ is proportional to the dynamic  tidal deformability. The universal relation is also broken if we switch from $\lambda_2^{\rm dyn}$ to $\lambda_2$.}
    \label{fig:Fr}
\end{figure*}

Notice that the energy flux can be computed using the field at infinity and horizon and the self force can be computed using the local field near the point mass. We can then subtract Eq.~\eqref{eq:Edottotal} and Eq.~\eqref{eq:Edottotalcs} to obtain the tide-induced flux difference and the tide-induced self-force difference. Their values should equal to each other according to the energy balance law, which can serve as a test of our calculation. The values obtained for the total tide-induced energy flux $\dot{E}^{\rm tide}_{\rm total}$  and tide-induced self force $F^{\rm t, tide}$ for the for the $l=2$, including both $m=\pm 2$ modes are listed in Table.~\ref{tab:comparison}. In this particular case, the equation of state of the central compact star is $p=\kappa\rho^{\Gamma}$, with $\Gamma = 2$, $\kappa = 107.32848\,{\rm km^2}$, mass $1.4M_{\odot}$. We can  find that the fractional differences between the values for various orbital frequencies are less than $2\times10^{-5}$, providing a confirmation of our numerical results. 


The radial component of the gravitational self force is conservative for a point mass following a circular motion we are considering. Similar to the evaluation of $F^{\rm t,tide}$, we can similarly compute $F^{\rm r, tide}$ using the local fields.
The numerical values for the $l=2$, including both $m=\pm 2$ modes are shown in Fig.~\ref{fig:Ft} and Fig.~\ref{fig:Fr}. We can  find that tide-induced gravitational self force $F^{\rm r, tide}$  follows a rather linear relation to the dynamic tidal deformability for the dynamical range considered. On the other hand, there is a visible nonlinear trend for  $F^{t,\rm tide}$, which shows up for sufficiently large $\lambda_2^{\rm dyn}$. As discussed in the flux section, this is probably due to the excitation of other modes of the star. Nevertheless, both $F^{\rm t,tide}$ and $F^{\rm r, tide}$ show rather decent universal behaviour with the dynamic tidal deformability. Notice that if we use the equilibrium  tidal deformability instead of the dynamic tidal deformability, the universal relations are clearly broken as shown in the right panels of Fig.~\ref{fig:Ft} and Fig.~\ref{fig:Fr}.

Therefore similar to the previous section, we can establish  universal relations between the tide-induced gravitational self-force $F^{\rm t,tide}$, $F^{\rm r, tide}$ and the dynamic tidal deformability $\lambda_2^{\rm dyn}$. The fitting formula for $F^{\rm t, tide}$ can be derived straightforwardly from the formula for the tide-induced energy flux at infinity as discussed in the previous section. Consequently we only need to find a fitting formula for $F^{\rm r,tide}$ for the $l=2$, including both $m=\pm 2$ modes.

\begin{figure} 
    \centering
    \includegraphics[height=6.0cm,width=8.7cm]{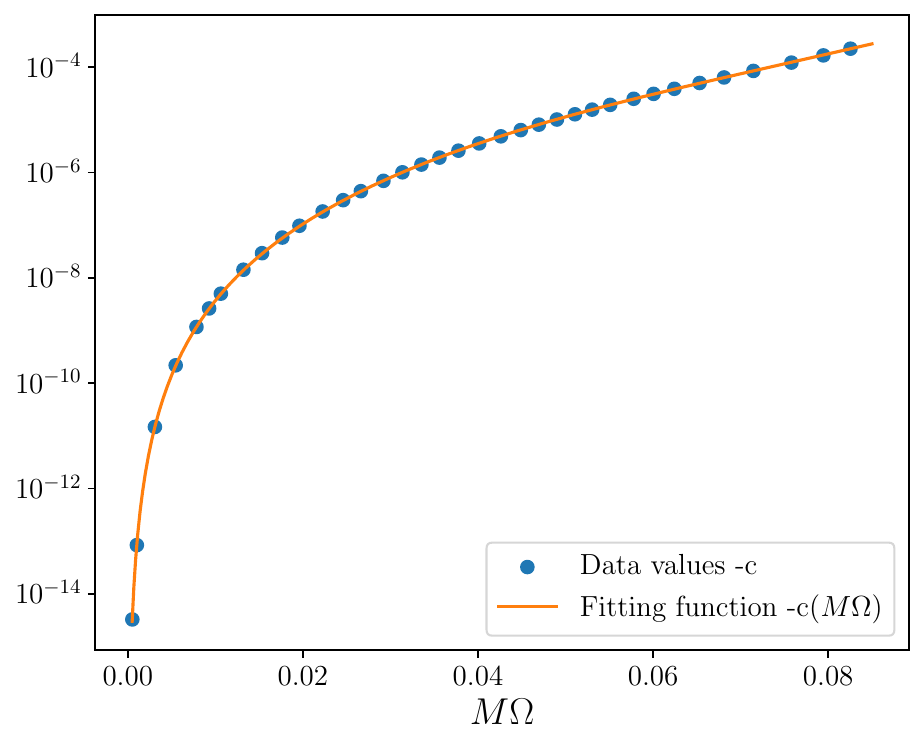}
    \caption{Data values $-c$ and the corresponding fitting functions $-c(M\Omega)$ are shown. We use several compact stars with polytropic equations of state $n=0.6$ to generate the numerical data. 
    }
    \label{fig:cfig}
\end{figure}

\begin{figure} 
    \centering
    \includegraphics[height=6.0cm,width=8.7cm]{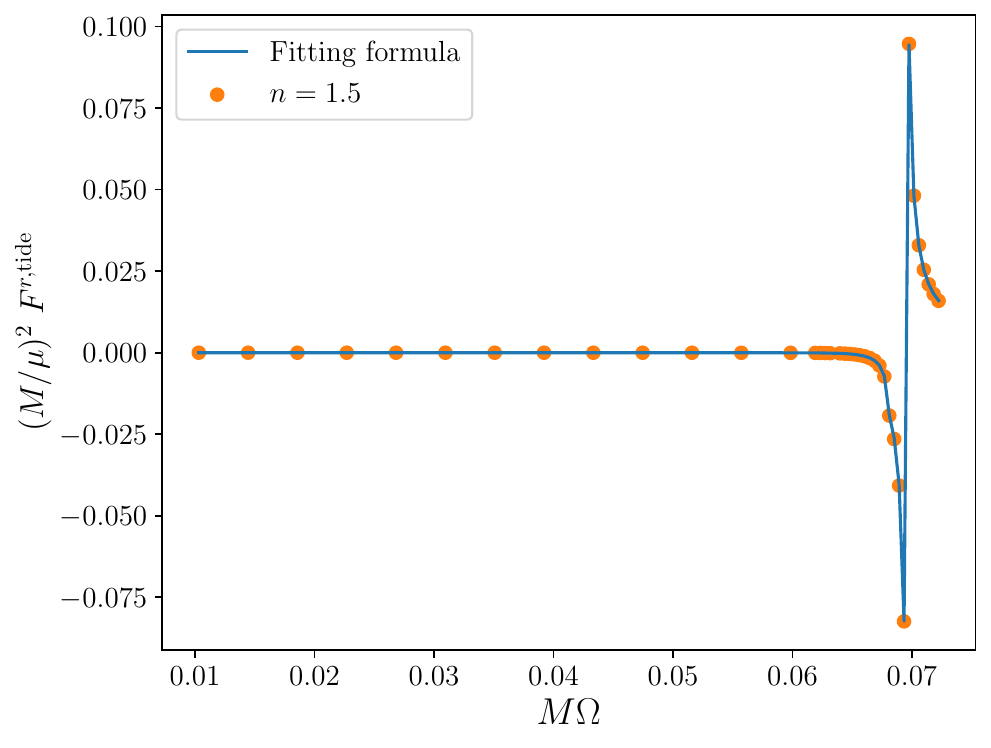}
    \caption{Comparison between our fitting formula Eq.~\eqref{eq:fitting_fr} of tide-induced self force and that produced by a point particle orbiting around a neutron star with polytropic equation of state $n=1.5$, $\rho_{c}=5\times10^{15}\,{\rm g/cm^3}$ and $\kappa=12.445425 \,{\rm km^2} $. In the most range of inspiral process, the error is less than $5\%$ unless very near the f-mode frequency which we do not show in this figure.}
    \label{fig:comparison_c}
\end{figure}

We assume the following form for the fitting formula of  $F^{\rm r,tide}$  by

\begin{align}\label{eq:fitting_fr}
    F^{\rm r,tide}=\lambda_2^{\rm dyn} c(M\Omega)\,,
\end{align}
where by fitting with numerical data $c(M\Omega)$ is well approximated by (see Fig.~\ref{fig:cfig})
\begin{align}
    c(M\Omega)&=(M\Omega)^{14/3}\big(-7.6023-3.6672\time 10^{3}(M\Omega)^2 \nonumber\\
    &\quad+ 1.1071\time 10 ^{6}(M\Omega)^4
    -1.1531\times 10^{7}(M\Omega)^5\big)\,.
\end{align}
Similar to the comparison of tide-induced gravitational flux $P^{\rm tide}$ in the last section, we compare the tide-induced self force $F^{\rm r, tide}$ obtained by our fitting formula Eq.~\eqref{eq:fitting_fr}, and that produced by a point particle orbiting around a neutron star with polytropic equation of state $n=1.5$. The detailed parameter of the star and comparison results are shown in Fig.~\ref{fig:comparison_c}. Our analysis reveals that, across the majority of the inspiral process, the discrepancy remains below $5\%$, except in regions very close to the f-mode frequency.

\subsection{Higher order contributions}\label{sec:ho}

At this point, it is also instructive to also check high-order contributions in the spherical modes decomposition. 
Just like $k_2$ describes the tidal response of the star to the quadrupole tidal field, $k_\ell$ (with $\ell \ge 3$) is needed to describe the response to higher-order tidal fields, which is necessary to obtain a complete understanding of the star's deformation. On the other hand, in the waveform construction perspective, it  is advantageous  to include fewer modes to reduce the number of parameters. 
In this section, we first numerically check how well the $\ell=2$ component represent the total flux, i.e., the $F^t$ component of the self force. In other words, we compare the total flux with only $\ell =2$ modes included and all $\ell \le 6$ modes included. 
Notice that the orbital frequency that resonantly
excite  the $\ell\geq3$ f-mode for   is lower than that needed to excite the  $\ell=2$ f-mode. Therefore for a certain orbital frequency, the tide-induced energy flux $P^{\rm tide}$ at infinity for $\ell=2$ is positive, while $P^{\rm tide}$ at infinity for $\ell \geq3$ may be either positive or negative. The higher-order contributions do not always increase the tide-induced energy flux. 

\begin{figure} 
    \centering
    \includegraphics[height=5.9cm,width=8.7cm]{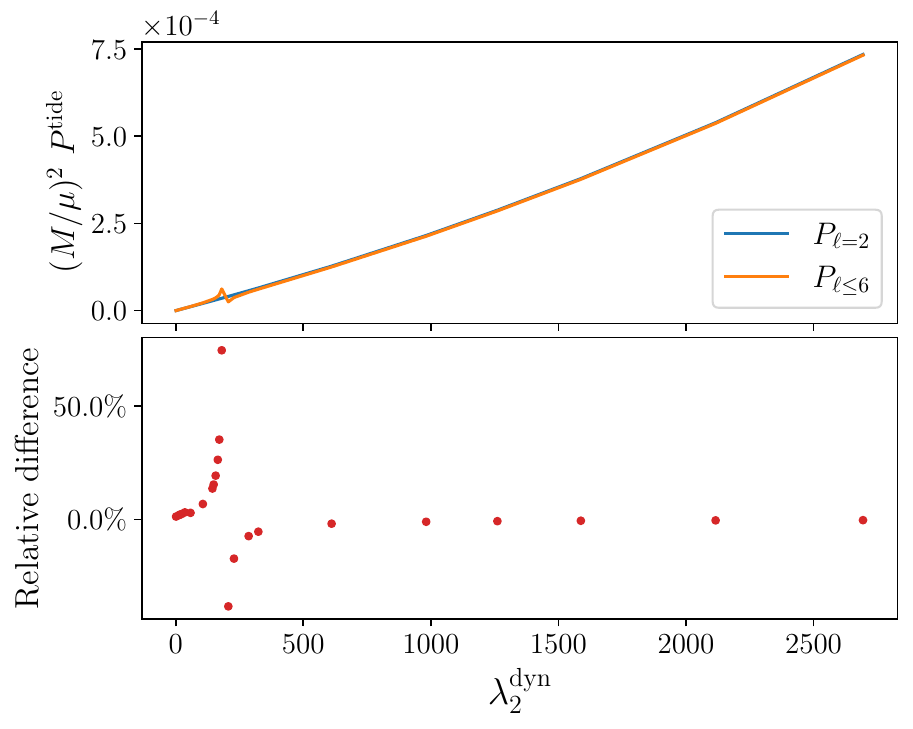}
    \caption{Higher order contributions of tide-induced gravitational flux $P^{\rm tide}$ at infinity at orbital frequency $\Omega=0.025$. In the majority of cases regarding dynamical tidal deformability, it is evident that higher-order $\ell\geq3$ contributions have minimal impact on the value of the $\ell=2$ energy flux, resulting in a variation of less than $3\%$. However, for certain values of the dynamical tidal deformability $\lambda_2^{\rm dyn}$, notable contributions are observed for $\ell\geq3$. In the plots, there are several peaks corresponds to the excitation of the f-mode for $\ell=3,4,5\, {\rm and}\, 6$ respectively, although we can not distinguish them accurately.}
    \label{fig:highorder1}
\end{figure}

\begin{figure} 
    \centering    \includegraphics[height=6.0cm,width=8.7cm]{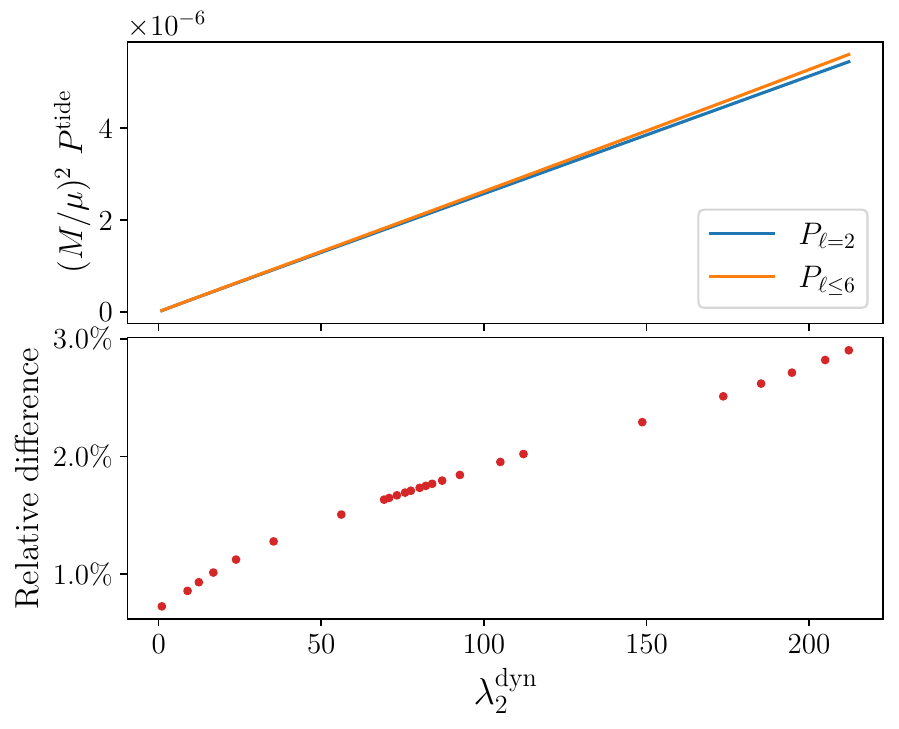}
    \caption{Higher order contributions of tide-induced gravitational flux at infinity at orbital frequency $M\Omega=0.0392$. In this case, it is evident that higher-order $\ell \geq3$ contributions have minimal impact on the value of the $\ell=2$ energy flux, resulting in a variation of less than $3\%$. Here $\ell=2,3,4$ f-mode resonances are not excited, except for $\ell=5,6$.}
    \label{fig:highorder2}
\end{figure}

In the Fig.~\ref{fig:highorder1} and Fig.~\ref{fig:highorder2}, we show higher-order contributions of tide-induced gravitational flux at infinity at two different orbital frequency $M\Omega=0.0516$ and $M\Omega=0.0392$. We choose a series of  star configurations as in the last section. In the case of $M\Omega=0.0392$, it is clear that higher-order contributions with $\ell \geq3$ have a negligible impact on the value of the energy flux comparing with the flux for $\ell=2$, resulting in a variation of less than $3\%$. Furthermore, it is noteworthy that f-modes  with $\ell=2,3,4$ are not resonantly excited in this case, while for certain  star configurations, f-modes with  $\ell=5,6$ are resonantly excited but the impact on the flux is small. 

On the other hand, for  $M\Omega=0.0516$, we still find that in most range of the dynamical tidal deformability, higher-order contributions have negligible effects on the value of the total energy flux compared with the $\ell=2$ flux, contributing to less than about $3\%$ variation. However, it is also observed that  for some certain values of $\lambda_2^{\rm dyn}$, significant flux variations are present due to the $\ell=3$ resonance. 
Therefore, the universal relation may only be a reasonable approximation away from the resonances and the resonance crossing may also be a necessary ingredient in constructing the waveform for a point mass moving around a compact star, in addition to the self-force resonance \cite{Flanagan:2010cd} and/or the tidal resonance \cite{Bonga:2019ycj}.

\begin{figure*} 
    \centering    \includegraphics[height=6.3cm,width=17cm]{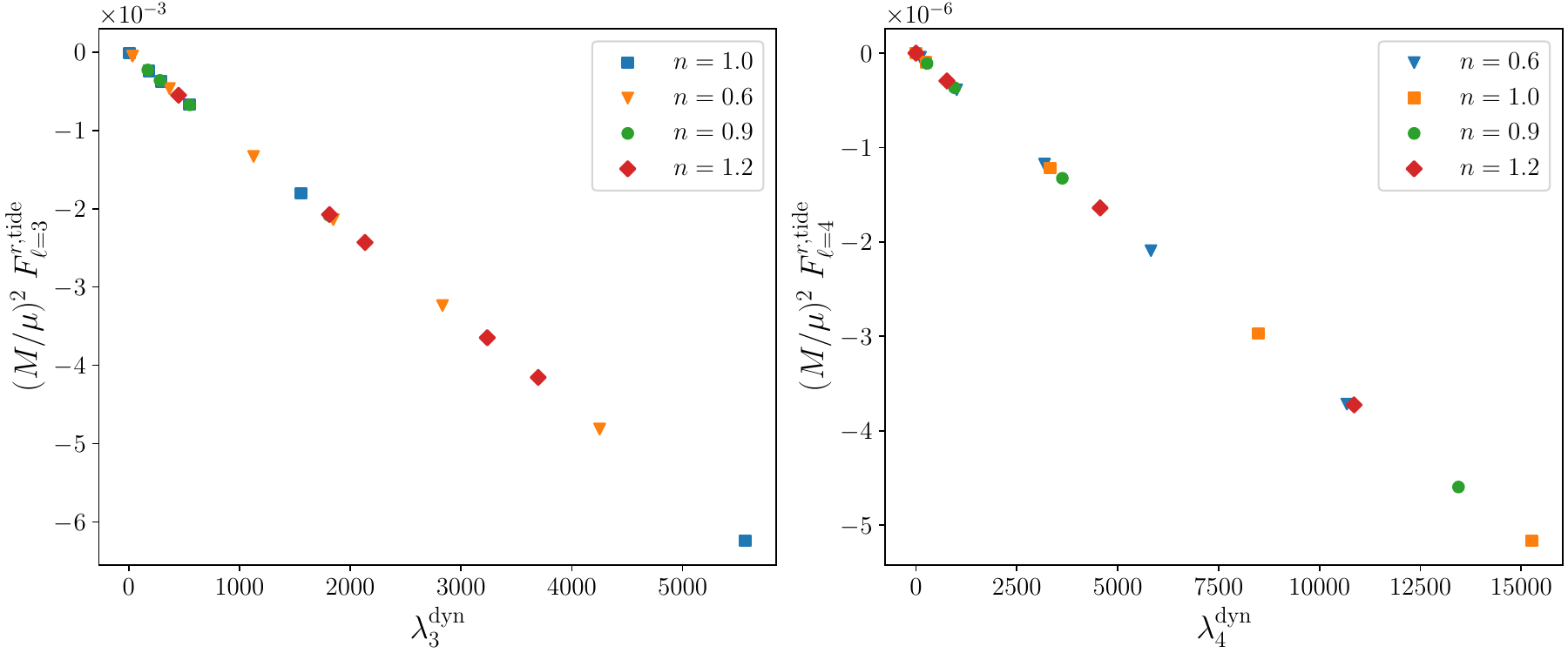}
    \caption{Higher-order contributions of the tide-induced gravitational self-force $F^{r,\rm tide}$ for the $\ell=3$ and $\ell=4$ modes. In the left panel, all modes with $|m|\leq3$ are included, with an orbital frequency being $M\Omega=0.0468$. In the right panel, all modes with $|m|\leq 4$ are included, with an orbital frequency of $M\Omega=0.0206$. Notably, for both $\ell=3$ and $\ell=4$ mode, a linear universal relationship reasonably hold with respect to the $\ell$-mode dynamic tidal deformability.}
    \label{fig:Fr_l3l4}
\end{figure*}

In fact, for modes with different  $\ell$, the impact of resonance is quite different. J. A. Pons \cite{pons2002gravitational} have shown that the peaks corresponding to higher $\ell$ modes display narrower widths relative to the orbital frequency, but have higher value. Indeed we also find that the resonance peaks for $\ell>4$ are numerically difficult to locate. Nevertheless transient crossing through resonances associated with higher order modes should matter less than that with lower-order modes.

For the radial component of the self force $F^r$, the higher $\ell$ component is not necessarily negligible. In this case, one may need to extend the universal description to include higher-order tidal deformations:
\begin{align}
F^{\rm r,tide} = \sum_{\ell \ge 2} \lambda^{\rm dyn}_\ell c_\ell (\Omega)
\end{align}
if nonlinearity in $\lambda^{\rm dyn}_\ell$ remains small. Here $\lambda^{\rm dyn}_\ell$ is defined as
\begin{align}
    \lambda_{\ell}^{\rm dyn}=\lambda_{\ell}\frac{\omega_{\ell,f}^2}{\omega_{\ell,f}^2-\omega^2}\,.
\end{align}
where $\omega_{\ell,f}$ is the f-mode frequency of $\ell$ mode, which can be approximately obtained in \cite{chan2014multipolar}. The dimensionless tidal deformability $\lambda_\ell$ is related to the tidal deformability as $\lambda_\ell = \tilde{\lambda}_\ell M^{-\ell}$, which relates the $\ell$-order multiple moment $Q_{\langle a_1,..,a_\ell \rangle}$ and tidal tensor $E_{\langle a_1,..,a_\ell\rangle}$
\begin{align}
Q_{\langle a_1,..,a_\ell \rangle} = \lambda_\ell E_{\langle a_1,..,a_\ell \rangle}\,,
\end{align}
where $\langle ... \rangle$ is the symmetric and trace-free operation for the indices. 

Indeed we compute $F^{r,\rm tide}$ for higher multipoles and show its dependence on $\lambda^{\rm dyn}_\ell$ for various equation of state in Fig.~\ref{fig:Fr_l3l4}. For the case of $\ell=3,4$ and the parameter range investigated here, linearity in $\lambda^{\rm dyn}_\ell$ remains a decent approximation and we  find that the universal relation still hold, i.e., points from different equation of state still lie on the same line. It is also interesting to check the relation with additional star equation of state and higher $\ell$.

\subsection{Tidal correction to the orbital phase}\label{sec:tct}
We are now ready to compute the tidal correction to the orbital phase, which is directly related to the waveform phase. The orbital phase, as a function of the orbital frequency is given by 
\begin{align}\label{eq:orbitalphase}
	\frac{d \phi}{d\Omega}&=\Omega\frac{dE/d\Omega}{P}\,.
\end{align}
As we are interested in the tidal correction,  we perturb the right hand side of the above equation to the linear order in the tidal perturbation in energy $E^{\rm tide}$ and tidal perturbation to the flux $P^{\rm tide}$:
\begin{align}\label{eq:dphitide}
\frac{d \phi^{\rm tide}}{d \Omega} =  \Omega\left ( \frac{dE^{\rm tide}/d\Omega}{P^{\rm pm}} -P^{\rm tide} \frac{dE^{\rm pm}/d\Omega}{(P^{\rm pm})^2} \right )\,,
\end{align}  
Here $E^{\rm pm}, P^{\rm pm}$ correspond to the limit that the tidal Love number is zero, i.e., the limit that satisfies by a black hole. However, since the compact star considered here is not dissipative and it does not have a horizon, the horizon flux component is not included in $P^{\rm pm}$. 

In the mass ratio expansion perspective, the first term on the right hand side is proportional to $\lambda \mathcal{O}(M/\mu)^0$ and the second term is proportional to $\lambda \mathcal{O}(M/\mu)^1$. Therefore in the extreme mass-ratio limit, the second term dominates over the first term in the orbital and gravitational phase. Therefore we only consider the second term as the leading order tidal correction to the phase. It is also worth to note that, if this is result is extended to the comparable mass ratio binaries for characterizing the tidal effects in the strong-gravity regime, the sub-leading terms in the mass ratio may also be important to compute, including the sub-leading order term  in $P^{\rm tide}$.

\begin{figure} 
    \centering
    \includegraphics[height=6.4cm,width=9cm]{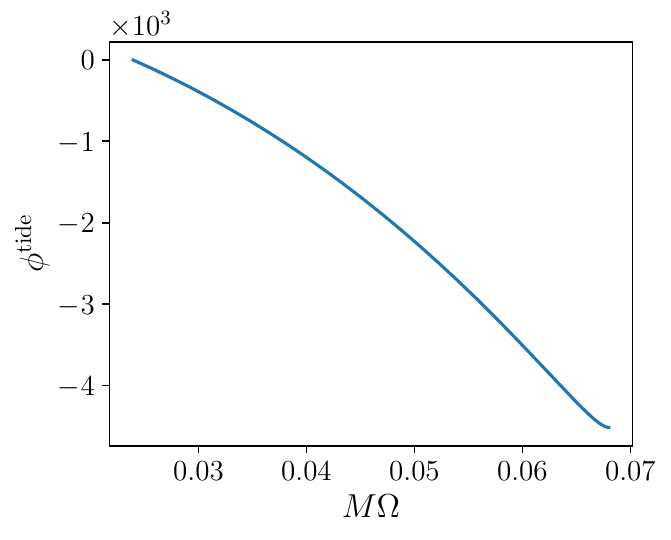}
    \caption{Tidal phase correction to the waveform of the EMRI system which has masses $(M,\mu)=(10^6,10)M_{\odot}$. This range of orbital frequency corresponds to the four-year inspiral evolution of an EMRI stopping at the  frequency the Inner Most Stable Orbit ($6M$). The central objecet is a supermassive compact object which can be considered as a black hole mimicker.}
    \label{fig:tidalphase}
\end{figure}

To illustrate the tidal correction to the phase, we consider a system corresponding to the four-year inspiral evolution of an EMRI, plunging into the central object at the end of the observation. The system has component masses $(M,\mu)=(10^6,10)M_{\odot}$. The EMRI starts from a certain orbital frequency and ends at the location of ISCO(Innermost Stable Circular Orbits), assuming a four-year observation time. 
 The initial  orbital frequency is about $M\Omega=0.0239$.
 The central object is chosen as supermassive compact with a central density $\rho_{c}=1.4\times 10^{6}\,{\rm g/cm^3}$, $\kappa=1.608\times10^9\,{\rm km^2}$ and $n=1.5$. This set of parameter gives rise to a star mass at about $10^6 M_\odot$ and the radius at about $8\times 10^6$\,km, which is smaller than the radius of ISCO. We plot the tidal orbital phase correction as a function in Fig.~\ref{fig:tidalphase}, using Eq.~\eqref{eq:dphitide}. Notice that at the end of the evolution, the tide-induced phase shift has already reached the level of $\sim 4.5 \times 10^3$ rads. Although this phase shift is a function of all the system parameter assumed, it already illustrates the importance of including the tide-induced phase shift in the construction of waveforms for black hole mimickers. Fortunately with the universal relation between the flux and the dynamic tidal deformability, this effect can be efficiently characterized by the equilibrium tidal deformability and the f-mode frequency. 

\section{conclusion}

It is known that for a static vacuum spacetime, the asymptotic multipole moments defined at spatial infinity uniquely determine the spacetime metric in the vacuum regime \cite{Geroch:1970cd}. Given the moments, the explicit reconstruction procedure for the metric is discussed in \cite{Fodor:2020fnq}, which has been used to compute the corresponding EMRI waveform in \cite{Tahura:2023qqt}. In the dynamic setting, it is also possible that a similar uniqueness theorem also exists for the radiative moments, which indicates that the particle experiences the same dissipative and conservative self force given the same radiative moments, regardless of the nature of the central engine. In the case that the radiative moments and the deformation of the central object are mainly contributed by a single mode, which is true for the particle$+$star scenario studied in this work, the gravitational self-force exerted on the particle can be efficiently described by a few parameters related to the tidal deformability of the star, without the need of specifying the structure of the central object. It will be interesting to extend the analysis presented in this work for other types of compact objects, i.e. boson stars \cite{Vincent:2015xta,Olivares:2018abq,Rosa:2023qcv,Cardoso:2019rvt}, wormholes \cite{Tsukamoto:2012xs,Tsukamoto:2016zdu,Mazza:2021rgq,Poisson:1995sv}, gravastars/AdS bubbles \cite{Pani:2009ss,Danielsson:2021ykm,Yang:2022gic,Mazur:2001fv}, etc., and check the validity of the universal description of the gravitational self force (away from mode resonances). It will also be important to extend the parameter regime of the orbit, e.g. including  eccentric orbits.

We have shown the tidal deformation of the central object may greatly impact the EMRI waveform through the modification of the gravitational wave flux. Therefore it is necessary to incorporate the finite size effect in the searching waveform models for black hole mimickers using EMRI observation. On the other hand, if an universal description holds for most of the compact black hole mimickers, the corresponding search waveform will be generally applicable and computationally efficient, as only a few more parameters are required to characterize the finite size effects. The treatment of resonant mode crossing will be similar to those for the transient self-force resonance \cite{Flanagan:2010cd} and tidal resonance \cite{Yang:2017aht,Bonga:2019ycj}, where a finite ``jump" of conserved quantities are expected across the resonances.

In \cite{Feng:2021sax} we have used a black hole perturbation perspective to study the tidal effect of comparable mass-ratio binaries containing neutron stars. The advantage of this approach is that it naturally captures the tidal effects in the strong gravity regime without performing Post-Newtonian expansions, but it requires a separate expansion in the mass ratio. In \cite{Feng:2021sax} we have argued that two model problems are useful for building the tidal waveform using black hole perturbation theory. The first model problem is to consider a compact star moving in a black hole spacetime, as already studied in \cite{Feng:2021sax}. The second model problem is to consider a point mass orbiting around a star, as considered in this work. In a future work we will discuss how to utilize these two calculations to build better tidal waveform for double neutron star and black hole-neutron star binaries, with comparisons to numerical simulations. 

\acknowledgements

We thank Eric Poisson for useful discussions.  H. Y. was supported by the Natural Sciences and
Engineering Research Council of Canada and in part by
Perimeter Institute for Theoretical Physics.
 Research at Perimeter Institute is supported in part by the Government of Canada through the Department of Innovation, Science and Economic Development Canada and by the Province of Ontario through the Ministry of Colleges and Universities. 

\appendix
\section{Source terms}\label{sec:st}
We consider the perturbing matter to be a particle of mass $\mu(\leq M)$ moving on a geodesic. \cite{zerilli1970gravitational}
\begin{align}
        S_{\rm odd}&=8\pi\left\{\frac{e^{-2\lambda}}{\sqrt{n+1}}Q+\frac{re^{-\lambda}}{\sqrt{2n(n+1)}}\frac{d(e^{-\lambda}D)}{dr}\right\}\,,\\
	S_{\rm even}&=8\pi e^{-\lambda}\biggl\{\frac{d}{dr}\left[\frac{r^2e^{-\lambda}}{\omega(nr+3M)}\left(\frac{A^{(1)}}{\sqrt{2}}+\frac{B^{(0)}}{\sqrt{n+1}}\right)\right]\nonumber\\
	&\quad-\frac{nr^2e^{-\lambda}A^{(1)}}{\sqrt{2}\omega(nr+3M)^2}+\frac{r^2e^{-\lambda}}{nr+3M}\left(A+\frac{B}{\sqrt{n+1}}\right)\nonumber\\
 &\quad-\frac{n(n+1)r^2+3nMr+6M^2}{\sqrt{n+1}\omega(nr+3M)^2}B^{(0)}\nonumber\\
	&\quad-\frac{2r}{\sqrt{2n(n+1)}}F\biggr\}\,,
\end{align}
where 
\begin{align}
        Q&=\frac{\mu\gamma e^{\lambda}}{r\sqrt{n+1}}\delta(r-R(t))\frac{dR}{dt}\bigg[\frac{1}{\sin\Theta}\frac{\partial Y^*_{\ell m}}{\partial\varphi}\frac{d\Theta}{dt}\nonumber\\
        &\qquad-\sin\Theta\frac{\partial Y^*_{\ell m}}{\partial\theta}\frac{d\Phi}{dt}\bigg]\,,\\
        D&=-\frac{\mu\gamma\delta(r-R(t))}{\sqrt{2n(n+1)}}\bigg\{\frac{1}{2}\left[\left(\frac{d\Theta}{dt}\right)^2-\sin^2\left(\frac{d\Phi}{dt}\right)^2\right]\nonumber\\
        &\quad\times\frac{1}{\sin\Theta}X^*_{\ell m}-\sin\Theta\frac{d\Phi}{dt}\frac{d\Theta}{dt}W^*_{\ell m}\bigg\}\,,\\
	A&=\mu \gamma\left(\frac{dR}{dt}\right)^2\frac{e^{2\lambda}}{r^2}\delta(r-R(t))Y_{\ell m}^{*}(\Omega)\,,\\
	A^{(1)}&=\sqrt{2}i\mu\gamma\frac{dR}{dt}r^{-2}\delta(r-R(t))Y_{\ell m}^{*}(\Omega)\,,
\end{align}
\begin{align}
	B^{(0)}&=\frac{i\mu}{\sqrt{n+1}}\gamma r^{-1}e^{-\lambda}\delta(r-R(t))\frac{d Y_{\ell m}^{*}}{dt}\,,\\
	B&=\frac{\mu}{\sqrt{n+1}}\gamma r^{-1}e^{\lambda}\frac{dR}{dt}\delta(r-R(t))\frac{d Y_{\ell m}^{*}}{dt}\,,\\
	F&=\frac{\mu}{\sqrt{2n(n+1)}}\gamma\delta(r-R(t))\bigg\{\frac{d\Theta}{dt}\frac{d\Phi}{dt}X_{\ell m}^{*}(\Omega)\nonumber\\
	&+\frac{1}{2}\left[\left(\frac{d\Theta}{dt}\right)^2-(\sin\Theta)^2\left(\frac{d\Phi}{dt}\right)\right]W_{lm}^{*}(\Omega)\bigg\}\,,
\end{align}
with
\begin{align}
	\Omega&=(\Theta,\Phi),\quad\gamma=\frac{dT}{dt}\,,\\
    X_{\ell m}&=2\frac{\partial}{\partial\varphi}\left(\frac{\partial}{\partial\theta}-\cot\theta\right)Y_{\ell m}\,,\\
    W_{\ell m}&=\left(\frac{\partial^2}{\partial\theta^2}-\cot\theta\frac{\partial}{\partial\theta}-\frac{1}{\sin^2\theta}\frac{\partial^2}{\partial\varphi^2}\right)Y_{\ell m}\,.
\end{align}

\bibliography{refs.bib}

\begin{thebibliography}{63}%
\makeatletter
\providecommand \@ifxundefined [1]{%
 \@ifx{#1\undefined}
}%
\providecommand \@ifnum [1]{%
 \ifnum #1\expandafter \@firstoftwo
 \else \expandafter \@secondoftwo
 \fi
}%
\providecommand \@ifx [1]{%
 \ifx #1\expandafter \@firstoftwo
 \else \expandafter \@secondoftwo
 \fi
}%
\providecommand \natexlab [1]{#1}%
\providecommand \enquote  [1]{``#1''}%
\providecommand \bibnamefont  [1]{#1}%
\providecommand \bibfnamefont [1]{#1}%
\providecommand \citenamefont [1]{#1}%
\providecommand \href@noop [0]{\@secondoftwo}%
\providecommand \href [0]{\begingroup \@sanitize@url \@href}%
\providecommand \@href[1]{\@@startlink{#1}\@@href}%
\providecommand \@@href[1]{\endgroup#1\@@endlink}%
\providecommand \@sanitize@url [0]{\catcode `\\12\catcode `\$12\catcode
  `\&12\catcode `\#12\catcode `\^12\catcode `\_12\catcode `\%12\relax}%
\providecommand \@@startlink[1]{}%
\providecommand \@@endlink[0]{}%
\providecommand \url  [0]{\begingroup\@sanitize@url \@url }%
\providecommand \@url [1]{\endgroup\@href {#1}{\urlprefix }}%
\providecommand \urlprefix  [0]{URL }%
\providecommand \Eprint [0]{\href }%
\providecommand \doibase [0]{http://dx.doi.org/}%
\providecommand \selectlanguage [0]{\@gobble}%
\providecommand \bibinfo  [0]{\@secondoftwo}%
\providecommand \bibfield  [0]{\@secondoftwo}%
\providecommand \translation [1]{[#1]}%
\providecommand \BibitemOpen [0]{}%
\providecommand \bibitemStop [0]{}%
\providecommand \bibitemNoStop [0]{.\EOS\space}%
\providecommand \EOS [0]{\spacefactor3000\relax}%
\providecommand \BibitemShut  [1]{\csname bibitem#1\endcsname}%
\let\auto@bib@innerbib\@empty
\bibitem [{\citenamefont {Baker}\ \emph {et~al.}(2019)\citenamefont {Baker}
  \emph {et~al.}}]{Baker:2019nia}%
  \BibitemOpen
  \bibfield  {author} {\bibinfo {author} {\bibfnamefont {J.}~\bibnamefont
  {Baker}} \emph {et~al.},\ }\href@noop {} {\  (\bibinfo {year} {2019})},\
  \Eprint {http://arxiv.org/abs/1907.06482} {arXiv:1907.06482 [astro-ph.IM]}
  \BibitemShut {NoStop}%
\bibitem [{\citenamefont {Mei}\ \emph {et~al.}(2021)\citenamefont {Mei} \emph
  {et~al.}}]{TianQin:2020hid}%
  \BibitemOpen
  \bibfield  {author} {\bibinfo {author} {\bibfnamefont {J.}~\bibnamefont
  {Mei}} \emph {et~al.} (\bibinfo {collaboration} {TianQin}),\ }\href {\doibase
  10.1093/ptep/ptaa114} {\bibfield  {journal} {\bibinfo  {journal} {PTEP}\
  }\textbf {\bibinfo {volume} {2021}},\ \bibinfo {pages} {05A107} (\bibinfo
  {year} {2021})},\ \Eprint {http://arxiv.org/abs/2008.10332} {arXiv:2008.10332
  [gr-qc]} \BibitemShut {NoStop}%
\bibitem [{\citenamefont {Hu}\ and\ \citenamefont {Wu}(2017)}]{Hu:2017mde}%
  \BibitemOpen
  \bibfield  {author} {\bibinfo {author} {\bibfnamefont {W.-R.}\ \bibnamefont
  {Hu}}\ and\ \bibinfo {author} {\bibfnamefont {Y.-L.}\ \bibnamefont {Wu}},\
  }\href {\doibase 10.1093/nsr/nwx116} {\bibfield  {journal} {\bibinfo
  {journal} {Natl. Sci. Rev.}\ }\textbf {\bibinfo {volume} {4}},\ \bibinfo
  {pages} {685} (\bibinfo {year} {2017})}\BibitemShut {NoStop}%
\bibitem [{\citenamefont {Babak}\ \emph {et~al.}(2017)\citenamefont {Babak},
  \citenamefont {Gair}, \citenamefont {Sesana}, \citenamefont {Barausse},
  \citenamefont {Sopuerta}, \citenamefont {Berry}, \citenamefont {Berti},
  \citenamefont {Amaro-Seoane}, \citenamefont {Petiteau},\ and\ \citenamefont
  {Klein}}]{Babak:2017tow}%
  \BibitemOpen
  \bibfield  {author} {\bibinfo {author} {\bibfnamefont {S.}~\bibnamefont
  {Babak}}, \bibinfo {author} {\bibfnamefont {J.}~\bibnamefont {Gair}},
  \bibinfo {author} {\bibfnamefont {A.}~\bibnamefont {Sesana}}, \bibinfo
  {author} {\bibfnamefont {E.}~\bibnamefont {Barausse}}, \bibinfo {author}
  {\bibfnamefont {C.~F.}\ \bibnamefont {Sopuerta}}, \bibinfo {author}
  {\bibfnamefont {C.~P.~L.}\ \bibnamefont {Berry}}, \bibinfo {author}
  {\bibfnamefont {E.}~\bibnamefont {Berti}}, \bibinfo {author} {\bibfnamefont
  {P.}~\bibnamefont {Amaro-Seoane}}, \bibinfo {author} {\bibfnamefont
  {A.}~\bibnamefont {Petiteau}}, \ and\ \bibinfo {author} {\bibfnamefont
  {A.}~\bibnamefont {Klein}},\ }\href {\doibase 10.1103/PhysRevD.95.103012}
  {\bibfield  {journal} {\bibinfo  {journal} {Phys. Rev. D}\ }\textbf {\bibinfo
  {volume} {95}},\ \bibinfo {pages} {103012} (\bibinfo {year} {2017})},\
  \Eprint {http://arxiv.org/abs/1703.09722} {arXiv:1703.09722 [gr-qc]}
  \BibitemShut {NoStop}%
\bibitem [{\citenamefont {Levin}(2007)}]{Levin:2006uc}%
  \BibitemOpen
  \bibfield  {author} {\bibinfo {author} {\bibfnamefont {Y.}~\bibnamefont
  {Levin}},\ }\href {\doibase 10.1111/j.1365-2966.2006.11155.x} {\bibfield
  {journal} {\bibinfo  {journal} {Mon. Not. Roy. Astron. Soc.}\ }\textbf
  {\bibinfo {volume} {374}},\ \bibinfo {pages} {515} (\bibinfo {year}
  {2007})},\ \Eprint {http://arxiv.org/abs/astro-ph/0603583}
  {arXiv:astro-ph/0603583} \BibitemShut {NoStop}%
\bibitem [{\citenamefont {Pan}\ and\ \citenamefont
  {Yang}(2021{\natexlab{a}})}]{Pan:2021ksp}%
  \BibitemOpen
  \bibfield  {author} {\bibinfo {author} {\bibfnamefont {Z.}~\bibnamefont
  {Pan}}\ and\ \bibinfo {author} {\bibfnamefont {H.}~\bibnamefont {Yang}},\
  }\href {\doibase 10.1103/PhysRevD.103.103018} {\bibfield  {journal} {\bibinfo
   {journal} {Phys. Rev. D}\ }\textbf {\bibinfo {volume} {103}},\ \bibinfo
  {pages} {103018} (\bibinfo {year} {2021}{\natexlab{a}})},\ \Eprint
  {http://arxiv.org/abs/2101.09146} {arXiv:2101.09146 [astro-ph.HE]}
  \BibitemShut {NoStop}%
\bibitem [{\citenamefont {Pan}\ \emph {et~al.}(2021)\citenamefont {Pan},
  \citenamefont {Lyu},\ and\ \citenamefont {Yang}}]{Pan:2021oob}%
  \BibitemOpen
  \bibfield  {author} {\bibinfo {author} {\bibfnamefont {Z.}~\bibnamefont
  {Pan}}, \bibinfo {author} {\bibfnamefont {Z.}~\bibnamefont {Lyu}}, \ and\
  \bibinfo {author} {\bibfnamefont {H.}~\bibnamefont {Yang}},\ }\href {\doibase
  10.1103/PhysRevD.104.063007} {\bibfield  {journal} {\bibinfo  {journal}
  {Phys. Rev. D}\ }\textbf {\bibinfo {volume} {104}},\ \bibinfo {pages}
  {063007} (\bibinfo {year} {2021})},\ \Eprint
  {http://arxiv.org/abs/2104.01208} {arXiv:2104.01208 [astro-ph.HE]}
  \BibitemShut {NoStop}%
\bibitem [{\citenamefont {Naoz}\ \emph {et~al.}(2022)\citenamefont {Naoz},
  \citenamefont {Rose}, \citenamefont {Michaely}, \citenamefont {Melchor},
  \citenamefont {Ramirez-Ruiz}, \citenamefont {Mockler},\ and\ \citenamefont
  {Schnittman}}]{Naoz:2022rru}%
  \BibitemOpen
  \bibfield  {author} {\bibinfo {author} {\bibfnamefont {S.}~\bibnamefont
  {Naoz}}, \bibinfo {author} {\bibfnamefont {S.~C.}\ \bibnamefont {Rose}},
  \bibinfo {author} {\bibfnamefont {E.}~\bibnamefont {Michaely}}, \bibinfo
  {author} {\bibfnamefont {D.}~\bibnamefont {Melchor}}, \bibinfo {author}
  {\bibfnamefont {E.}~\bibnamefont {Ramirez-Ruiz}}, \bibinfo {author}
  {\bibfnamefont {B.}~\bibnamefont {Mockler}}, \ and\ \bibinfo {author}
  {\bibfnamefont {J.~D.}\ \bibnamefont {Schnittman}},\ }\href {\doibase
  10.3847/2041-8213/ac574b} {\bibfield  {journal} {\bibinfo  {journal}
  {Astrophys. J. Lett.}\ }\textbf {\bibinfo {volume} {927}},\ \bibinfo {pages}
  {L18} (\bibinfo {year} {2022})},\ \Eprint {http://arxiv.org/abs/2202.12303}
  {arXiv:2202.12303 [astro-ph.HE]} \BibitemShut {NoStop}%
\bibitem [{\citenamefont {Bode}\ and\ \citenamefont
  {Wegg}(2014)}]{Bode:2013mma}%
  \BibitemOpen
  \bibfield  {author} {\bibinfo {author} {\bibfnamefont {J.~N.}\ \bibnamefont
  {Bode}}\ and\ \bibinfo {author} {\bibfnamefont {C.}~\bibnamefont {Wegg}},\
  }\href {\doibase 10.1093/mnras/stt2227} {\bibfield  {journal} {\bibinfo
  {journal} {Mon. Not. Roy. Astron. Soc.}\ }\textbf {\bibinfo {volume} {438}},\
  \bibinfo {pages} {573} (\bibinfo {year} {2014})},\ \Eprint
  {http://arxiv.org/abs/1310.5745} {arXiv:1310.5745 [astro-ph.CO]} \BibitemShut
  {NoStop}%
\bibitem [{\citenamefont {Mazzolari}\ \emph {et~al.}(2022)\citenamefont
  {Mazzolari}, \citenamefont {Bonetti}, \citenamefont {Sesana}, \citenamefont
  {Colombo}, \citenamefont {Dotti}, \citenamefont {Lodato},\ and\ \citenamefont
  {Izquierdo-Villalba}}]{Mazzolari:2022cho}%
  \BibitemOpen
  \bibfield  {author} {\bibinfo {author} {\bibfnamefont {G.}~\bibnamefont
  {Mazzolari}}, \bibinfo {author} {\bibfnamefont {M.}~\bibnamefont {Bonetti}},
  \bibinfo {author} {\bibfnamefont {A.}~\bibnamefont {Sesana}}, \bibinfo
  {author} {\bibfnamefont {R.~M.}\ \bibnamefont {Colombo}}, \bibinfo {author}
  {\bibfnamefont {M.}~\bibnamefont {Dotti}}, \bibinfo {author} {\bibfnamefont
  {G.}~\bibnamefont {Lodato}}, \ and\ \bibinfo {author} {\bibfnamefont
  {D.}~\bibnamefont {Izquierdo-Villalba}},\ }\href {\doibase
  10.1093/mnras/stac2255} {\bibfield  {journal} {\bibinfo  {journal} {Mon. Not.
  Roy. Astron. Soc.}\ }\textbf {\bibinfo {volume} {516}},\ \bibinfo {pages}
  {1959} (\bibinfo {year} {2022})},\ \Eprint {http://arxiv.org/abs/2204.05343}
  {arXiv:2204.05343 [astro-ph.GA]} \BibitemShut {NoStop}%
\bibitem [{\citenamefont {Pan}\ and\ \citenamefont
  {Yang}(2021{\natexlab{b}})}]{Pan:2021xhv}%
  \BibitemOpen
  \bibfield  {author} {\bibinfo {author} {\bibfnamefont {Z.}~\bibnamefont
  {Pan}}\ and\ \bibinfo {author} {\bibfnamefont {H.}~\bibnamefont {Yang}},\
  }\href {\doibase 10.3847/1538-4357/ac249c} {\bibfield  {journal} {\bibinfo
  {journal} {Astrophys. J.}\ }\textbf {\bibinfo {volume} {923}},\ \bibinfo
  {pages} {173} (\bibinfo {year} {2021}{\natexlab{b}})},\ \Eprint
  {http://arxiv.org/abs/2108.00267} {arXiv:2108.00267 [astro-ph.HE]}
  \BibitemShut {NoStop}%
\bibitem [{\citenamefont {Bonga}\ \emph {et~al.}(2019)\citenamefont {Bonga},
  \citenamefont {Yang},\ and\ \citenamefont {Hughes}}]{Bonga:2019ycj}%
  \BibitemOpen
  \bibfield  {author} {\bibinfo {author} {\bibfnamefont {B.}~\bibnamefont
  {Bonga}}, \bibinfo {author} {\bibfnamefont {H.}~\bibnamefont {Yang}}, \ and\
  \bibinfo {author} {\bibfnamefont {S.~A.}\ \bibnamefont {Hughes}},\ }\href
  {\doibase 10.1103/PhysRevLett.123.101103} {\bibfield  {journal} {\bibinfo
  {journal} {Phys. Rev. Lett.}\ }\textbf {\bibinfo {volume} {123}},\ \bibinfo
  {pages} {101103} (\bibinfo {year} {2019})},\ \Eprint
  {http://arxiv.org/abs/1905.00030} {arXiv:1905.00030 [gr-qc]} \BibitemShut
  {NoStop}%
\bibitem [{\citenamefont {Yunes}\ \emph {et~al.}(2011)\citenamefont {Yunes},
  \citenamefont {Kocsis}, \citenamefont {Loeb},\ and\ \citenamefont
  {Haiman}}]{Yunes:2011ws}%
  \BibitemOpen
  \bibfield  {author} {\bibinfo {author} {\bibfnamefont {N.}~\bibnamefont
  {Yunes}}, \bibinfo {author} {\bibfnamefont {B.}~\bibnamefont {Kocsis}},
  \bibinfo {author} {\bibfnamefont {A.}~\bibnamefont {Loeb}}, \ and\ \bibinfo
  {author} {\bibfnamefont {Z.}~\bibnamefont {Haiman}},\ }\href {\doibase
  10.1103/PhysRevLett.107.171103} {\bibfield  {journal} {\bibinfo  {journal}
  {Phys. Rev. Lett.}\ }\textbf {\bibinfo {volume} {107}},\ \bibinfo {pages}
  {171103} (\bibinfo {year} {2011})},\ \Eprint {http://arxiv.org/abs/1103.4609}
  {arXiv:1103.4609 [astro-ph.CO]} \BibitemShut {NoStop}%
\bibitem [{\citenamefont {Barausse}\ \emph {et~al.}(2015)\citenamefont
  {Barausse}, \citenamefont {Cardoso},\ and\ \citenamefont
  {Pani}}]{Barausse:2014pra}%
  \BibitemOpen
  \bibfield  {author} {\bibinfo {author} {\bibfnamefont {E.}~\bibnamefont
  {Barausse}}, \bibinfo {author} {\bibfnamefont {V.}~\bibnamefont {Cardoso}}, \
  and\ \bibinfo {author} {\bibfnamefont {P.}~\bibnamefont {Pani}},\ }\href
  {\doibase 10.1088/1742-6596/610/1/012044} {\bibfield  {journal} {\bibinfo
  {journal} {J. Phys. Conf. Ser.}\ }\textbf {\bibinfo {volume} {610}},\
  \bibinfo {pages} {012044} (\bibinfo {year} {2015})},\ \Eprint
  {http://arxiv.org/abs/1404.7140} {arXiv:1404.7140 [astro-ph.CO]} \BibitemShut
  {NoStop}%
\bibitem [{\citenamefont {Zhang}\ and\ \citenamefont
  {Yang}(2020)}]{Zhang:2019eid}%
  \BibitemOpen
  \bibfield  {author} {\bibinfo {author} {\bibfnamefont {J.}~\bibnamefont
  {Zhang}}\ and\ \bibinfo {author} {\bibfnamefont {H.}~\bibnamefont {Yang}},\
  }\href {\doibase 10.1103/PhysRevD.101.043020} {\bibfield  {journal} {\bibinfo
   {journal} {Phys. Rev. D}\ }\textbf {\bibinfo {volume} {101}},\ \bibinfo
  {pages} {043020} (\bibinfo {year} {2020})},\ \Eprint
  {http://arxiv.org/abs/1907.13582} {arXiv:1907.13582 [gr-qc]} \BibitemShut
  {NoStop}%
\bibitem [{\citenamefont {Zhang}\ and\ \citenamefont
  {Yang}(2019)}]{Zhang:2018kib}%
  \BibitemOpen
  \bibfield  {author} {\bibinfo {author} {\bibfnamefont {J.}~\bibnamefont
  {Zhang}}\ and\ \bibinfo {author} {\bibfnamefont {H.}~\bibnamefont {Yang}},\
  }\href {\doibase 10.1103/PhysRevD.99.064018} {\bibfield  {journal} {\bibinfo
  {journal} {Phys. Rev. D}\ }\textbf {\bibinfo {volume} {99}},\ \bibinfo
  {pages} {064018} (\bibinfo {year} {2019})},\ \Eprint
  {http://arxiv.org/abs/1808.02905} {arXiv:1808.02905 [gr-qc]} \BibitemShut
  {NoStop}%
\bibitem [{\citenamefont {Hannuksela}\ \emph {et~al.}(2019)\citenamefont
  {Hannuksela}, \citenamefont {Wong}, \citenamefont {Brito}, \citenamefont
  {Berti},\ and\ \citenamefont {Li}}]{Hannuksela:2018izj}%
  \BibitemOpen
  \bibfield  {author} {\bibinfo {author} {\bibfnamefont {O.~A.}\ \bibnamefont
  {Hannuksela}}, \bibinfo {author} {\bibfnamefont {K.~W.~K.}\ \bibnamefont
  {Wong}}, \bibinfo {author} {\bibfnamefont {R.}~\bibnamefont {Brito}},
  \bibinfo {author} {\bibfnamefont {E.}~\bibnamefont {Berti}}, \ and\ \bibinfo
  {author} {\bibfnamefont {T.~G.~F.}\ \bibnamefont {Li}},\ }\href {\doibase
  10.1038/s41550-019-0712-4} {\bibfield  {journal} {\bibinfo  {journal} {Nature
  Astron.}\ }\textbf {\bibinfo {volume} {3}},\ \bibinfo {pages} {447} (\bibinfo
  {year} {2019})},\ \Eprint {http://arxiv.org/abs/1804.09659} {arXiv:1804.09659
  [astro-ph.HE]} \BibitemShut {NoStop}%
\bibitem [{\citenamefont {Ryan}(1995)}]{Ryan:1995wh}%
  \BibitemOpen
  \bibfield  {author} {\bibinfo {author} {\bibfnamefont {F.~D.}\ \bibnamefont
  {Ryan}},\ }\href {\doibase 10.1103/PhysRevD.52.5707} {\bibfield  {journal}
  {\bibinfo  {journal} {Phys. Rev. D}\ }\textbf {\bibinfo {volume} {52}},\
  \bibinfo {pages} {5707} (\bibinfo {year} {1995})}\BibitemShut {NoStop}%
\bibitem [{\citenamefont {Raposo}\ \emph {et~al.}(2019)\citenamefont {Raposo},
  \citenamefont {Pani},\ and\ \citenamefont {Emparan}}]{Raposo:2018xkf}%
  \BibitemOpen
  \bibfield  {author} {\bibinfo {author} {\bibfnamefont {G.}~\bibnamefont
  {Raposo}}, \bibinfo {author} {\bibfnamefont {P.}~\bibnamefont {Pani}}, \ and\
  \bibinfo {author} {\bibfnamefont {R.}~\bibnamefont {Emparan}},\ }\href
  {\doibase 10.1103/PhysRevD.99.104050} {\bibfield  {journal} {\bibinfo
  {journal} {Phys. Rev. D}\ }\textbf {\bibinfo {volume} {99}},\ \bibinfo
  {pages} {104050} (\bibinfo {year} {2019})},\ \Eprint
  {http://arxiv.org/abs/1812.07615} {arXiv:1812.07615 [gr-qc]} \BibitemShut
  {NoStop}%
\bibitem [{\citenamefont {Tahura}\ \emph {et~al.}(2023)\citenamefont {Tahura},
  \citenamefont {Khalvati},\ and\ \citenamefont {Yang}}]{Tahura:2023qqt}%
  \BibitemOpen
  \bibfield  {author} {\bibinfo {author} {\bibfnamefont {S.}~\bibnamefont
  {Tahura}}, \bibinfo {author} {\bibfnamefont {H.}~\bibnamefont {Khalvati}}, \
  and\ \bibinfo {author} {\bibfnamefont {H.}~\bibnamefont {Yang}},\ }\href@noop
  {} {\  (\bibinfo {year} {2023})},\ \Eprint {http://arxiv.org/abs/2309.11491}
  {arXiv:2309.11491 [gr-qc]} \BibitemShut {NoStop}%
\bibitem [{\citenamefont {Cardoso}\ and\ \citenamefont
  {Pani}(2019)}]{Cardoso:2019rvt}%
  \BibitemOpen
  \bibfield  {author} {\bibinfo {author} {\bibfnamefont {V.}~\bibnamefont
  {Cardoso}}\ and\ \bibinfo {author} {\bibfnamefont {P.}~\bibnamefont {Pani}},\
  }\href {\doibase 10.1007/s41114-019-0020-4} {\bibfield  {journal} {\bibinfo
  {journal} {Living Rev. Rel.}\ }\textbf {\bibinfo {volume} {22}},\ \bibinfo
  {pages} {4} (\bibinfo {year} {2019})},\ \Eprint
  {http://arxiv.org/abs/1904.05363} {arXiv:1904.05363 [gr-qc]} \BibitemShut
  {NoStop}%
\bibitem [{\citenamefont {Barausse}\ \emph {et~al.}(2020)\citenamefont
  {Barausse} \emph {et~al.}}]{Barausse:2020rsu}%
  \BibitemOpen
  \bibfield  {author} {\bibinfo {author} {\bibfnamefont {E.}~\bibnamefont
  {Barausse}} \emph {et~al.},\ }\href {\doibase 10.1007/s10714-020-02691-1}
  {\bibfield  {journal} {\bibinfo  {journal} {Gen. Rel. Grav.}\ }\textbf
  {\bibinfo {volume} {52}},\ \bibinfo {pages} {81} (\bibinfo {year} {2020})},\
  \Eprint {http://arxiv.org/abs/2001.09793} {arXiv:2001.09793 [gr-qc]}
  \BibitemShut {NoStop}%
\bibitem [{\citenamefont {Arun}\ \emph {et~al.}(2022)\citenamefont {Arun} \emph
  {et~al.}}]{LISA:2022kgy}%
  \BibitemOpen
  \bibfield  {author} {\bibinfo {author} {\bibfnamefont {K.~G.}\ \bibnamefont
  {Arun}} \emph {et~al.} (\bibinfo {collaboration} {LISA}),\ }\href {\doibase
  10.1007/s41114-022-00036-9} {\bibfield  {journal} {\bibinfo  {journal}
  {Living Rev. Rel.}\ }\textbf {\bibinfo {volume} {25}},\ \bibinfo {pages} {4}
  (\bibinfo {year} {2022})},\ \Eprint {http://arxiv.org/abs/2205.01597}
  {arXiv:2205.01597 [gr-qc]} \BibitemShut {NoStop}%
\bibitem [{\citenamefont {Afshordi}\ \emph {et~al.}(2023)\citenamefont
  {Afshordi} \emph {et~al.}}]{LISAConsortiumWaveformWorkingGroup:2023arg}%
  \BibitemOpen
  \bibfield  {author} {\bibinfo {author} {\bibfnamefont {N.}~\bibnamefont
  {Afshordi}} \emph {et~al.} (\bibinfo {collaboration} {LISA Consortium
  Waveform Working Group}),\ }\href@noop {} {\  (\bibinfo {year} {2023})},\
  \Eprint {http://arxiv.org/abs/2311.01300} {arXiv:2311.01300 [gr-qc]}
  \BibitemShut {NoStop}%
\bibitem [{\citenamefont {Maggio}\ \emph
  {et~al.}(2021{\natexlab{a}})\citenamefont {Maggio}, \citenamefont {van~de
  Meent},\ and\ \citenamefont {Pani}}]{Maggio:2021uge}%
  \BibitemOpen
  \bibfield  {author} {\bibinfo {author} {\bibfnamefont {E.}~\bibnamefont
  {Maggio}}, \bibinfo {author} {\bibfnamefont {M.}~\bibnamefont {van~de
  Meent}}, \ and\ \bibinfo {author} {\bibfnamefont {P.}~\bibnamefont {Pani}},\
  }\href {\doibase 10.1103/PhysRevD.104.104026} {\bibfield  {journal} {\bibinfo
   {journal} {Phys. Rev. D}\ }\textbf {\bibinfo {volume} {104}},\ \bibinfo
  {pages} {104026} (\bibinfo {year} {2021}{\natexlab{a}})},\ \Eprint
  {http://arxiv.org/abs/2106.07195} {arXiv:2106.07195 [gr-qc]} \BibitemShut
  {NoStop}%
\bibitem [{\citenamefont {Thorne}\ and\ \citenamefont
  {Campolattaro}(1967)}]{thorne1967non}%
  \BibitemOpen
  \bibfield  {author} {\bibinfo {author} {\bibfnamefont {K.~S.}\ \bibnamefont
  {Thorne}}\ and\ \bibinfo {author} {\bibfnamefont {A.}~\bibnamefont
  {Campolattaro}},\ }\href@noop {} {\bibfield  {journal} {\bibinfo  {journal}
  {The astrophysical journal}\ }\textbf {\bibinfo {volume} {149}},\ \bibinfo
  {pages} {591} (\bibinfo {year} {1967})}\BibitemShut {NoStop}%
\bibitem [{\citenamefont {Lindblom}\ and\ \citenamefont
  {Detweiler}(1983)}]{lindblom1983quadrupole}%
  \BibitemOpen
  \bibfield  {author} {\bibinfo {author} {\bibfnamefont {L.}~\bibnamefont
  {Lindblom}}\ and\ \bibinfo {author} {\bibfnamefont {S.~L.}\ \bibnamefont
  {Detweiler}},\ }\href@noop {} {\bibfield  {journal} {\bibinfo  {journal} {The
  Astrophysical Journal Supplement Series}\ }\textbf {\bibinfo {volume} {53}},\
  \bibinfo {pages} {73} (\bibinfo {year} {1983})}\BibitemShut {NoStop}%
\bibitem [{\citenamefont {Kojima}(1987)}]{kojima1987stellar}%
  \BibitemOpen
  \bibfield  {author} {\bibinfo {author} {\bibfnamefont {Y.}~\bibnamefont
  {Kojima}},\ }\href@noop {} {\bibfield  {journal} {\bibinfo  {journal}
  {Progress of theoretical physics}\ }\textbf {\bibinfo {volume} {77}},\
  \bibinfo {pages} {297} (\bibinfo {year} {1987})}\BibitemShut {NoStop}%
\bibitem [{\citenamefont {Tominaga}\ \emph {et~al.}(1999)\citenamefont
  {Tominaga}, \citenamefont {Saijo},\ and\ \citenamefont
  {Maeda}}]{tominaga1999gravitational}%
  \BibitemOpen
  \bibfield  {author} {\bibinfo {author} {\bibfnamefont {K.}~\bibnamefont
  {Tominaga}}, \bibinfo {author} {\bibfnamefont {M.}~\bibnamefont {Saijo}}, \
  and\ \bibinfo {author} {\bibfnamefont {K.-i.}\ \bibnamefont {Maeda}},\
  }\href@noop {} {\bibfield  {journal} {\bibinfo  {journal} {Physical Review
  D}\ }\textbf {\bibinfo {volume} {60}},\ \bibinfo {pages} {024004} (\bibinfo
  {year} {1999})}\BibitemShut {NoStop}%
\bibitem [{\citenamefont {Gittins}\ \emph {et~al.}(2020)\citenamefont
  {Gittins}, \citenamefont {Andersson},\ and\ \citenamefont
  {Pereira}}]{gittins2020tidal}%
  \BibitemOpen
  \bibfield  {author} {\bibinfo {author} {\bibfnamefont {F.}~\bibnamefont
  {Gittins}}, \bibinfo {author} {\bibfnamefont {N.}~\bibnamefont {Andersson}},
  \ and\ \bibinfo {author} {\bibfnamefont {J.~P.}\ \bibnamefont {Pereira}},\
  }\href@noop {} {\bibfield  {journal} {\bibinfo  {journal} {Physical Review
  D}\ }\textbf {\bibinfo {volume} {101}},\ \bibinfo {pages} {103025} (\bibinfo
  {year} {2020})}\BibitemShut {NoStop}%
\bibitem [{\citenamefont {Datta}\ \emph {et~al.}(2020)\citenamefont {Datta},
  \citenamefont {Brito}, \citenamefont {Bose}, \citenamefont {Pani},\ and\
  \citenamefont {Hughes}}]{datta2020tidal}%
  \BibitemOpen
  \bibfield  {author} {\bibinfo {author} {\bibfnamefont {S.}~\bibnamefont
  {Datta}}, \bibinfo {author} {\bibfnamefont {R.}~\bibnamefont {Brito}},
  \bibinfo {author} {\bibfnamefont {S.}~\bibnamefont {Bose}}, \bibinfo {author}
  {\bibfnamefont {P.}~\bibnamefont {Pani}}, \ and\ \bibinfo {author}
  {\bibfnamefont {S.~A.}\ \bibnamefont {Hughes}},\ }\href@noop {} {\bibfield
  {journal} {\bibinfo  {journal} {Physical Review D}\ }\textbf {\bibinfo
  {volume} {101}},\ \bibinfo {pages} {044004} (\bibinfo {year}
  {2020})}\BibitemShut {NoStop}%
\bibitem [{\citenamefont {Fransen}\ \emph {et~al.}(2021)\citenamefont
  {Fransen}, \citenamefont {Tielemans}, \citenamefont {Vercnocke},\ and\
  \citenamefont {Koekoek}}]{fransen2021modeling}%
  \BibitemOpen
  \bibfield  {author} {\bibinfo {author} {\bibfnamefont {K.}~\bibnamefont
  {Fransen}}, \bibinfo {author} {\bibfnamefont {R.}~\bibnamefont {Tielemans}},
  \bibinfo {author} {\bibfnamefont {B.}~\bibnamefont {Vercnocke}}, \ and\
  \bibinfo {author} {\bibfnamefont {G.}~\bibnamefont {Koekoek}},\ }\href@noop
  {} {\bibfield  {journal} {\bibinfo  {journal} {Physical Review D}\ }\textbf
  {\bibinfo {volume} {104}},\ \bibinfo {pages} {044044} (\bibinfo {year}
  {2021})}\BibitemShut {NoStop}%
\bibitem [{\citenamefont {Maggio}\ \emph
  {et~al.}(2021{\natexlab{b}})\citenamefont {Maggio}, \citenamefont {van~de
  Meent},\ and\ \citenamefont {Pani}}]{maggio2021extreme}%
  \BibitemOpen
  \bibfield  {author} {\bibinfo {author} {\bibfnamefont {E.}~\bibnamefont
  {Maggio}}, \bibinfo {author} {\bibfnamefont {M.}~\bibnamefont {van~de
  Meent}}, \ and\ \bibinfo {author} {\bibfnamefont {P.}~\bibnamefont {Pani}},\
  }\href@noop {} {\bibfield  {journal} {\bibinfo  {journal} {Physical Review
  D}\ }\textbf {\bibinfo {volume} {104}},\ \bibinfo {pages} {104026} (\bibinfo
  {year} {2021}{\natexlab{b}})}\BibitemShut {NoStop}%
\bibitem [{\citenamefont {Oppenheimer}\ and\ \citenamefont
  {Volkoff}(1939)}]{oppenheimer1939massive}%
  \BibitemOpen
  \bibfield  {author} {\bibinfo {author} {\bibfnamefont {J.~R.}\ \bibnamefont
  {Oppenheimer}}\ and\ \bibinfo {author} {\bibfnamefont {G.~M.}\ \bibnamefont
  {Volkoff}},\ }\href@noop {} {\bibfield  {journal} {\bibinfo  {journal}
  {Physical Review}\ }\textbf {\bibinfo {volume} {55}},\ \bibinfo {pages} {374}
  (\bibinfo {year} {1939})}\BibitemShut {NoStop}%
\bibitem [{\citenamefont {Feng}\ \emph {et~al.}(2022)\citenamefont {Feng},
  \citenamefont {Lyu},\ and\ \citenamefont {Yang}}]{Feng:2021sax}%
  \BibitemOpen
  \bibfield  {author} {\bibinfo {author} {\bibfnamefont {X.}~\bibnamefont
  {Feng}}, \bibinfo {author} {\bibfnamefont {Z.}~\bibnamefont {Lyu}}, \ and\
  \bibinfo {author} {\bibfnamefont {H.}~\bibnamefont {Yang}},\ }\href {\doibase
  10.1103/PhysRevD.105.104043} {\bibfield  {journal} {\bibinfo  {journal}
  {Phys. Rev. D}\ }\textbf {\bibinfo {volume} {105}},\ \bibinfo {pages}
  {104043} (\bibinfo {year} {2022})},\ \Eprint
  {http://arxiv.org/abs/2104.11848} {arXiv:2104.11848 [gr-qc]} \BibitemShut
  {NoStop}%
\bibitem [{\citenamefont {Nakamura}\ \emph {et~al.}(1987)\citenamefont
  {Nakamura}, \citenamefont {Oohara},\ and\ \citenamefont
  {Kojima}}]{nakamura1987general}%
  \BibitemOpen
  \bibfield  {author} {\bibinfo {author} {\bibfnamefont {T.}~\bibnamefont
  {Nakamura}}, \bibinfo {author} {\bibfnamefont {K.}~\bibnamefont {Oohara}}, \
  and\ \bibinfo {author} {\bibfnamefont {Y.}~\bibnamefont {Kojima}},\
  }\href@noop {} {\bibfield  {journal} {\bibinfo  {journal} {Progress of
  Theoretical Physics Supplement}\ }\textbf {\bibinfo {volume} {90}},\ \bibinfo
  {pages} {1} (\bibinfo {year} {1987})}\BibitemShut {NoStop}%
\bibitem [{\citenamefont {Detweiler}\ and\ \citenamefont
  {Lindblom}(1985)}]{detweiler1985nonradial}%
  \BibitemOpen
  \bibfield  {author} {\bibinfo {author} {\bibfnamefont {S.}~\bibnamefont
  {Detweiler}}\ and\ \bibinfo {author} {\bibfnamefont {L.}~\bibnamefont
  {Lindblom}},\ }\href@noop {} {\bibfield  {journal} {\bibinfo  {journal}
  {Astrophysical Journal, Part 1 (ISSN 0004-637X), vol. 292, May 1, 1985, p.
  12-15.}\ }\textbf {\bibinfo {volume} {292}},\ \bibinfo {pages} {12} (\bibinfo
  {year} {1985})}\BibitemShut {NoStop}%
\bibitem [{\citenamefont {Arfken}\ \emph {et~al.}(2011)\citenamefont {Arfken},
  \citenamefont {Weber},\ and\ \citenamefont
  {Harris}}]{arfken2011mathematical}%
  \BibitemOpen
  \bibfield  {author} {\bibinfo {author} {\bibfnamefont {G.~B.}\ \bibnamefont
  {Arfken}}, \bibinfo {author} {\bibfnamefont {H.~J.}\ \bibnamefont {Weber}}, \
  and\ \bibinfo {author} {\bibfnamefont {F.~E.}\ \bibnamefont {Harris}},\
  }\href@noop {} {\emph {\bibinfo {title} {Mathematical methods for physicists:
  a comprehensive guide}}}\ (\bibinfo  {publisher} {Academic press},\ \bibinfo
  {year} {2011})\BibitemShut {NoStop}%
\bibitem [{\citenamefont {Barack}\ and\ \citenamefont
  {Sago}(2007)}]{barack2007gravitational}%
  \BibitemOpen
  \bibfield  {author} {\bibinfo {author} {\bibfnamefont {L.}~\bibnamefont
  {Barack}}\ and\ \bibinfo {author} {\bibfnamefont {N.}~\bibnamefont {Sago}},\
  }\href@noop {} {\bibfield  {journal} {\bibinfo  {journal} {Physical Review
  D}\ }\textbf {\bibinfo {volume} {75}},\ \bibinfo {pages} {064021} (\bibinfo
  {year} {2007})}\BibitemShut {NoStop}%
\bibitem [{\citenamefont {Bini}\ \emph {et~al.}(2012)\citenamefont {Bini},
  \citenamefont {Damour},\ and\ \citenamefont {Faye}}]{bini2012effective}%
  \BibitemOpen
  \bibfield  {author} {\bibinfo {author} {\bibfnamefont {D.}~\bibnamefont
  {Bini}}, \bibinfo {author} {\bibfnamefont {T.}~\bibnamefont {Damour}}, \ and\
  \bibinfo {author} {\bibfnamefont {G.}~\bibnamefont {Faye}},\ }\href@noop {}
  {\bibfield  {journal} {\bibinfo  {journal} {Physical Review D}\ }\textbf
  {\bibinfo {volume} {85}},\ \bibinfo {pages} {124034} (\bibinfo {year}
  {2012})}\BibitemShut {NoStop}%
\bibitem [{\citenamefont {Chakrabarti}\ \emph {et~al.}(2013)\citenamefont
  {Chakrabarti}, \citenamefont {Delsate},\ and\ \citenamefont
  {Steinhoff}}]{chakrabarti2013effective}%
  \BibitemOpen
  \bibfield  {author} {\bibinfo {author} {\bibfnamefont {S.}~\bibnamefont
  {Chakrabarti}}, \bibinfo {author} {\bibfnamefont {T.}~\bibnamefont
  {Delsate}}, \ and\ \bibinfo {author} {\bibfnamefont {J.}~\bibnamefont
  {Steinhoff}},\ }\href@noop {} {\bibfield  {journal} {\bibinfo  {journal}
  {Physical Review D}\ }\textbf {\bibinfo {volume} {88}},\ \bibinfo {pages}
  {084038} (\bibinfo {year} {2013})}\BibitemShut {NoStop}%
\bibitem [{\citenamefont {Chan}\ \emph {et~al.}(2014)\citenamefont {Chan},
  \citenamefont {Sham}, \citenamefont {Leung},\ and\ \citenamefont
  {Lin}}]{chan2014multipolar}%
  \BibitemOpen
  \bibfield  {author} {\bibinfo {author} {\bibfnamefont {T.}~\bibnamefont
  {Chan}}, \bibinfo {author} {\bibfnamefont {Y.-H.}\ \bibnamefont {Sham}},
  \bibinfo {author} {\bibfnamefont {P.}~\bibnamefont {Leung}}, \ and\ \bibinfo
  {author} {\bibfnamefont {L.-M.}\ \bibnamefont {Lin}},\ }\href@noop {}
  {\bibfield  {journal} {\bibinfo  {journal} {Physical Review D}\ }\textbf
  {\bibinfo {volume} {90}},\ \bibinfo {pages} {124023} (\bibinfo {year}
  {2014})}\BibitemShut {NoStop}%
\bibitem [{\citenamefont {Shashank}\ \emph {et~al.}(2023)\citenamefont
  {Shashank}, \citenamefont {Nouri},\ and\ \citenamefont
  {Gupta}}]{shashank2023f}%
  \BibitemOpen
  \bibfield  {author} {\bibinfo {author} {\bibfnamefont {S.}~\bibnamefont
  {Shashank}}, \bibinfo {author} {\bibfnamefont {F.~H.}\ \bibnamefont {Nouri}},
  \ and\ \bibinfo {author} {\bibfnamefont {A.}~\bibnamefont {Gupta}},\
  }\href@noop {} {\bibfield  {journal} {\bibinfo  {journal} {New Astronomy}\
  }\textbf {\bibinfo {volume} {104}},\ \bibinfo {pages} {102067} (\bibinfo
  {year} {2023})}\BibitemShut {NoStop}%
\bibitem [{\citenamefont {Pitre}\ and\ \citenamefont
  {Poisson}(2024)}]{pitre2024general}%
  \BibitemOpen
  \bibfield  {author} {\bibinfo {author} {\bibfnamefont {T.}~\bibnamefont
  {Pitre}}\ and\ \bibinfo {author} {\bibfnamefont {E.}~\bibnamefont
  {Poisson}},\ }\href@noop {} {\bibfield  {journal} {\bibinfo  {journal}
  {Physical Review D}\ }\textbf {\bibinfo {volume} {109}},\ \bibinfo {pages}
  {064004} (\bibinfo {year} {2024})}\BibitemShut {NoStop}%
\bibitem [{\citenamefont {Pound}\ \emph {et~al.}(2014)\citenamefont {Pound},
  \citenamefont {Merlin},\ and\ \citenamefont
  {Barack}}]{pound2014gravitational}%
  \BibitemOpen
  \bibfield  {author} {\bibinfo {author} {\bibfnamefont {A.}~\bibnamefont
  {Pound}}, \bibinfo {author} {\bibfnamefont {C.}~\bibnamefont {Merlin}}, \
  and\ \bibinfo {author} {\bibfnamefont {L.}~\bibnamefont {Barack}},\
  }\href@noop {} {\bibfield  {journal} {\bibinfo  {journal} {Physical Review
  D}\ }\textbf {\bibinfo {volume} {89}},\ \bibinfo {pages} {024009} (\bibinfo
  {year} {2014})}\BibitemShut {NoStop}%
\bibitem [{\citenamefont {Detweiler}\ and\ \citenamefont
  {Whiting}(2003)}]{detweiler2003self}%
  \BibitemOpen
  \bibfield  {author} {\bibinfo {author} {\bibfnamefont {S.}~\bibnamefont
  {Detweiler}}\ and\ \bibinfo {author} {\bibfnamefont {B.~F.}\ \bibnamefont
  {Whiting}},\ }\href@noop {} {\bibfield  {journal} {\bibinfo  {journal}
  {Physical Review D}\ }\textbf {\bibinfo {volume} {67}},\ \bibinfo {pages}
  {024025} (\bibinfo {year} {2003})}\BibitemShut {NoStop}%
\bibitem [{\citenamefont {Flanagan}\ and\ \citenamefont
  {Hinderer}(2012)}]{Flanagan:2010cd}%
  \BibitemOpen
  \bibfield  {author} {\bibinfo {author} {\bibfnamefont {E.~E.}\ \bibnamefont
  {Flanagan}}\ and\ \bibinfo {author} {\bibfnamefont {T.}~\bibnamefont
  {Hinderer}},\ }\href {\doibase 10.1103/PhysRevLett.109.071102} {\bibfield
  {journal} {\bibinfo  {journal} {Phys. Rev. Lett.}\ }\textbf {\bibinfo
  {volume} {109}},\ \bibinfo {pages} {071102} (\bibinfo {year} {2012})},\
  \Eprint {http://arxiv.org/abs/1009.4923} {arXiv:1009.4923 [gr-qc]}
  \BibitemShut {NoStop}%
\bibitem [{\citenamefont {Pons}\ \emph {et~al.}(2002)\citenamefont {Pons},
  \citenamefont {Berti}, \citenamefont {Gualtieri}, \citenamefont {Miniutti},\
  and\ \citenamefont {Ferrari}}]{pons2002gravitational}%
  \BibitemOpen
  \bibfield  {author} {\bibinfo {author} {\bibfnamefont {J.}~\bibnamefont
  {Pons}}, \bibinfo {author} {\bibfnamefont {E.}~\bibnamefont {Berti}},
  \bibinfo {author} {\bibfnamefont {L.}~\bibnamefont {Gualtieri}}, \bibinfo
  {author} {\bibfnamefont {G.}~\bibnamefont {Miniutti}}, \ and\ \bibinfo
  {author} {\bibfnamefont {V.}~\bibnamefont {Ferrari}},\ }\href@noop {}
  {\bibfield  {journal} {\bibinfo  {journal} {Physical Review D}\ }\textbf
  {\bibinfo {volume} {65}},\ \bibinfo {pages} {104021} (\bibinfo {year}
  {2002})}\BibitemShut {NoStop}%
\bibitem [{\citenamefont {Geroch}(1970)}]{Geroch:1970cd}%
  \BibitemOpen
  \bibfield  {author} {\bibinfo {author} {\bibfnamefont {R.~P.}\ \bibnamefont
  {Geroch}},\ }\href {\doibase 10.1063/1.1665427} {\bibfield  {journal}
  {\bibinfo  {journal} {J. Math. Phys.}\ }\textbf {\bibinfo {volume} {11}},\
  \bibinfo {pages} {2580} (\bibinfo {year} {1970})}\BibitemShut {NoStop}%
\bibitem [{\citenamefont {Fodor}\ \emph {et~al.}(2021)\citenamefont {Fodor},
  \citenamefont {Filho},\ and\ \citenamefont {Hartmann}}]{Fodor:2020fnq}%
  \BibitemOpen
  \bibfield  {author} {\bibinfo {author} {\bibfnamefont {G.}~\bibnamefont
  {Fodor}}, \bibinfo {author} {\bibfnamefont {E.~d. S.~C.}\ \bibnamefont
  {Filho}}, \ and\ \bibinfo {author} {\bibfnamefont {B.}~\bibnamefont
  {Hartmann}},\ }\href {\doibase 10.1103/PhysRevD.104.064012} {\bibfield
  {journal} {\bibinfo  {journal} {Phys. Rev. D}\ }\textbf {\bibinfo {volume}
  {104}},\ \bibinfo {pages} {064012} (\bibinfo {year} {2021})},\ \Eprint
  {http://arxiv.org/abs/2012.05548} {arXiv:2012.05548 [gr-qc]} \BibitemShut
  {NoStop}%
\bibitem [{\citenamefont {Vincent}\ \emph {et~al.}(2016)\citenamefont
  {Vincent}, \citenamefont {Meliani}, \citenamefont {Grandclement},
  \citenamefont {Gourgoulhon},\ and\ \citenamefont {Straub}}]{Vincent:2015xta}%
  \BibitemOpen
  \bibfield  {author} {\bibinfo {author} {\bibfnamefont {F.~H.}\ \bibnamefont
  {Vincent}}, \bibinfo {author} {\bibfnamefont {Z.}~\bibnamefont {Meliani}},
  \bibinfo {author} {\bibfnamefont {P.}~\bibnamefont {Grandclement}}, \bibinfo
  {author} {\bibfnamefont {E.}~\bibnamefont {Gourgoulhon}}, \ and\ \bibinfo
  {author} {\bibfnamefont {O.}~\bibnamefont {Straub}},\ }\href {\doibase
  10.1088/0264-9381/33/10/105015} {\bibfield  {journal} {\bibinfo  {journal}
  {Class. Quant. Grav.}\ }\textbf {\bibinfo {volume} {33}},\ \bibinfo {pages}
  {105015} (\bibinfo {year} {2016})},\ \Eprint
  {http://arxiv.org/abs/1510.04170} {arXiv:1510.04170 [gr-qc]} \BibitemShut
  {NoStop}%
\bibitem [{\citenamefont {Olivares}\ \emph {et~al.}(2020)\citenamefont
  {Olivares}, \citenamefont {Younsi}, \citenamefont {Fromm}, \citenamefont
  {De~Laurentis}, \citenamefont {Porth}, \citenamefont {Mizuno}, \citenamefont
  {Falcke}, \citenamefont {Kramer},\ and\ \citenamefont
  {Rezzolla}}]{Olivares:2018abq}%
  \BibitemOpen
  \bibfield  {author} {\bibinfo {author} {\bibfnamefont {H.}~\bibnamefont
  {Olivares}}, \bibinfo {author} {\bibfnamefont {Z.}~\bibnamefont {Younsi}},
  \bibinfo {author} {\bibfnamefont {C.~M.}\ \bibnamefont {Fromm}}, \bibinfo
  {author} {\bibfnamefont {M.}~\bibnamefont {De~Laurentis}}, \bibinfo {author}
  {\bibfnamefont {O.}~\bibnamefont {Porth}}, \bibinfo {author} {\bibfnamefont
  {Y.}~\bibnamefont {Mizuno}}, \bibinfo {author} {\bibfnamefont
  {H.}~\bibnamefont {Falcke}}, \bibinfo {author} {\bibfnamefont
  {M.}~\bibnamefont {Kramer}}, \ and\ \bibinfo {author} {\bibfnamefont
  {L.}~\bibnamefont {Rezzolla}},\ }\href {\doibase 10.1093/mnras/staa1878}
  {\bibfield  {journal} {\bibinfo  {journal} {Mon. Not. Roy. Astron. Soc.}\
  }\textbf {\bibinfo {volume} {497}},\ \bibinfo {pages} {521} (\bibinfo {year}
  {2020})},\ \Eprint {http://arxiv.org/abs/1809.08682} {arXiv:1809.08682
  [gr-qc]} \BibitemShut {NoStop}%
\bibitem [{\citenamefont {Rosa}\ \emph {et~al.}(2023)\citenamefont {Rosa},
  \citenamefont {Macedo},\ and\ \citenamefont {Rubiera-Garcia}}]{Rosa:2023qcv}%
  \BibitemOpen
  \bibfield  {author} {\bibinfo {author} {\bibfnamefont {J.~a.~L.}\
  \bibnamefont {Rosa}}, \bibinfo {author} {\bibfnamefont {C.~F.~B.}\
  \bibnamefont {Macedo}}, \ and\ \bibinfo {author} {\bibfnamefont
  {D.}~\bibnamefont {Rubiera-Garcia}},\ }\href@noop {} {\  (\bibinfo {year}
  {2023})},\ \Eprint {http://arxiv.org/abs/2303.17296} {arXiv:2303.17296
  [gr-qc]} \BibitemShut {NoStop}%
\bibitem [{\citenamefont {Tsukamoto}\ \emph {et~al.}(2012)\citenamefont
  {Tsukamoto}, \citenamefont {Harada},\ and\ \citenamefont
  {Yajima}}]{Tsukamoto:2012xs}%
  \BibitemOpen
  \bibfield  {author} {\bibinfo {author} {\bibfnamefont {N.}~\bibnamefont
  {Tsukamoto}}, \bibinfo {author} {\bibfnamefont {T.}~\bibnamefont {Harada}}, \
  and\ \bibinfo {author} {\bibfnamefont {K.}~\bibnamefont {Yajima}},\ }\href
  {\doibase 10.1103/PhysRevD.86.104062} {\bibfield  {journal} {\bibinfo
  {journal} {Phys. Rev. D}\ }\textbf {\bibinfo {volume} {86}},\ \bibinfo
  {pages} {104062} (\bibinfo {year} {2012})},\ \Eprint
  {http://arxiv.org/abs/1207.0047} {arXiv:1207.0047 [gr-qc]} \BibitemShut
  {NoStop}%
\bibitem [{\citenamefont {Tsukamoto}\ and\ \citenamefont
  {Harada}(2017)}]{Tsukamoto:2016zdu}%
  \BibitemOpen
  \bibfield  {author} {\bibinfo {author} {\bibfnamefont {N.}~\bibnamefont
  {Tsukamoto}}\ and\ \bibinfo {author} {\bibfnamefont {T.}~\bibnamefont
  {Harada}},\ }\href {\doibase 10.1103/PhysRevD.95.024030} {\bibfield
  {journal} {\bibinfo  {journal} {Phys. Rev. D}\ }\textbf {\bibinfo {volume}
  {95}},\ \bibinfo {pages} {024030} (\bibinfo {year} {2017})},\ \Eprint
  {http://arxiv.org/abs/1607.01120} {arXiv:1607.01120 [gr-qc]} \BibitemShut
  {NoStop}%
\bibitem [{\citenamefont {Mazza}\ \emph {et~al.}(2021)\citenamefont {Mazza},
  \citenamefont {Franzin},\ and\ \citenamefont {Liberati}}]{Mazza:2021rgq}%
  \BibitemOpen
  \bibfield  {author} {\bibinfo {author} {\bibfnamefont {J.}~\bibnamefont
  {Mazza}}, \bibinfo {author} {\bibfnamefont {E.}~\bibnamefont {Franzin}}, \
  and\ \bibinfo {author} {\bibfnamefont {S.}~\bibnamefont {Liberati}},\ }\href
  {\doibase 10.1088/1475-7516/2021/04/082} {\bibfield  {journal} {\bibinfo
  {journal} {JCAP}\ }\textbf {\bibinfo {volume} {04}},\ \bibinfo {pages} {082}
  (\bibinfo {year} {2021})},\ \Eprint {http://arxiv.org/abs/2102.01105}
  {arXiv:2102.01105 [gr-qc]} \BibitemShut {NoStop}%
\bibitem [{\citenamefont {Poisson}\ and\ \citenamefont
  {Visser}(1995)}]{Poisson:1995sv}%
  \BibitemOpen
  \bibfield  {author} {\bibinfo {author} {\bibfnamefont {E.}~\bibnamefont
  {Poisson}}\ and\ \bibinfo {author} {\bibfnamefont {M.}~\bibnamefont
  {Visser}},\ }\href {\doibase 10.1103/PhysRevD.52.7318} {\bibfield  {journal}
  {\bibinfo  {journal} {Phys. Rev. D}\ }\textbf {\bibinfo {volume} {52}},\
  \bibinfo {pages} {7318} (\bibinfo {year} {1995})},\ \Eprint
  {http://arxiv.org/abs/gr-qc/9506083} {arXiv:gr-qc/9506083} \BibitemShut
  {NoStop}%
\bibitem [{\citenamefont {Pani}\ \emph {et~al.}(2009)\citenamefont {Pani},
  \citenamefont {Berti}, \citenamefont {Cardoso}, \citenamefont {Chen},\ and\
  \citenamefont {Norte}}]{Pani:2009ss}%
  \BibitemOpen
  \bibfield  {author} {\bibinfo {author} {\bibfnamefont {P.}~\bibnamefont
  {Pani}}, \bibinfo {author} {\bibfnamefont {E.}~\bibnamefont {Berti}},
  \bibinfo {author} {\bibfnamefont {V.}~\bibnamefont {Cardoso}}, \bibinfo
  {author} {\bibfnamefont {Y.}~\bibnamefont {Chen}}, \ and\ \bibinfo {author}
  {\bibfnamefont {R.}~\bibnamefont {Norte}},\ }\href {\doibase
  10.1103/PhysRevD.80.124047} {\bibfield  {journal} {\bibinfo  {journal} {Phys.
  Rev. D}\ }\textbf {\bibinfo {volume} {80}},\ \bibinfo {pages} {124047}
  (\bibinfo {year} {2009})},\ \Eprint {http://arxiv.org/abs/0909.0287}
  {arXiv:0909.0287 [gr-qc]} \BibitemShut {NoStop}%
\bibitem [{\citenamefont {Danielsson}\ \emph {et~al.}(2021)\citenamefont
  {Danielsson}, \citenamefont {Lehner},\ and\ \citenamefont
  {Pretorius}}]{Danielsson:2021ykm}%
  \BibitemOpen
  \bibfield  {author} {\bibinfo {author} {\bibfnamefont {U.}~\bibnamefont
  {Danielsson}}, \bibinfo {author} {\bibfnamefont {L.}~\bibnamefont {Lehner}},
  \ and\ \bibinfo {author} {\bibfnamefont {F.}~\bibnamefont {Pretorius}},\
  }\href {\doibase 10.1103/PhysRevD.104.124011} {\bibfield  {journal} {\bibinfo
   {journal} {Phys. Rev. D}\ }\textbf {\bibinfo {volume} {104}},\ \bibinfo
  {pages} {124011} (\bibinfo {year} {2021})},\ \Eprint
  {http://arxiv.org/abs/2109.09814} {arXiv:2109.09814 [gr-qc]} \BibitemShut
  {NoStop}%
\bibitem [{\citenamefont {Yang}\ \emph {et~al.}(2023)\citenamefont {Yang},
  \citenamefont {Bonga},\ and\ \citenamefont {Pan}}]{Yang:2022gic}%
  \BibitemOpen
  \bibfield  {author} {\bibinfo {author} {\bibfnamefont {H.}~\bibnamefont
  {Yang}}, \bibinfo {author} {\bibfnamefont {B.}~\bibnamefont {Bonga}}, \ and\
  \bibinfo {author} {\bibfnamefont {Z.}~\bibnamefont {Pan}},\ }\href {\doibase
  10.1103/PhysRevLett.130.011402} {\bibfield  {journal} {\bibinfo  {journal}
  {Phys. Rev. Lett.}\ }\textbf {\bibinfo {volume} {130}},\ \bibinfo {pages}
  {011402} (\bibinfo {year} {2023})},\ \Eprint
  {http://arxiv.org/abs/2207.13754} {arXiv:2207.13754 [gr-qc]} \BibitemShut
  {NoStop}%
\bibitem [{\citenamefont {Mazur}\ and\ \citenamefont
  {Mottola}(2023)}]{Mazur:2001fv}%
  \BibitemOpen
  \bibfield  {author} {\bibinfo {author} {\bibfnamefont {P.~O.}\ \bibnamefont
  {Mazur}}\ and\ \bibinfo {author} {\bibfnamefont {E.}~\bibnamefont
  {Mottola}},\ }\href {\doibase 10.3390/universe9020088} {\bibfield  {journal}
  {\bibinfo  {journal} {Universe}\ }\textbf {\bibinfo {volume} {9}},\ \bibinfo
  {pages} {88} (\bibinfo {year} {2023})},\ \Eprint
  {http://arxiv.org/abs/gr-qc/0109035} {arXiv:gr-qc/0109035} \BibitemShut
  {NoStop}%
\bibitem [{\citenamefont {Yang}\ and\ \citenamefont
  {Casals}(2017)}]{Yang:2017aht}%
  \BibitemOpen
  \bibfield  {author} {\bibinfo {author} {\bibfnamefont {H.}~\bibnamefont
  {Yang}}\ and\ \bibinfo {author} {\bibfnamefont {M.}~\bibnamefont {Casals}},\
  }\href {\doibase 10.1103/PhysRevD.96.083015} {\bibfield  {journal} {\bibinfo
  {journal} {Phys. Rev. D}\ }\textbf {\bibinfo {volume} {96}},\ \bibinfo
  {pages} {083015} (\bibinfo {year} {2017})},\ \Eprint
  {http://arxiv.org/abs/1704.02022} {arXiv:1704.02022 [gr-qc]} \BibitemShut
  {NoStop}%
\bibitem [{\citenamefont {Zerilli}(1970)}]{zerilli1970gravitational}%
  \BibitemOpen
  \bibfield  {author} {\bibinfo {author} {\bibfnamefont {F.~J.}\ \bibnamefont
  {Zerilli}},\ }\href@noop {} {\bibfield  {journal} {\bibinfo  {journal}
  {Physical Review D}\ }\textbf {\bibinfo {volume} {2}},\ \bibinfo {pages}
  {2141} (\bibinfo {year} {1970})}\BibitemShut {NoStop}%
\end{thebibliography}%

\end{document}